\documentclass{article}
\usepackage{amsmath}[1996/11/01]
\usepackage{amssymb,amsthm,amsxtra}

\def\[#1\]{\begin{equation}#1\end{equation}}
\makeatletter
\def\beq{%
   \relax\ifmmode
      \@badmath
   \else
      \ifvmode
         \nointerlineskip
         \makebox[.6\linewidth]%
      \fi
      $$
   \fi
}
\def\eeq{%
   \relax\ifmmode
      \ifinner
         \@badmath
      \else
         $$
      \fi
   \else
      \@badmath
   \fi
   \ignorespaces
}

\def\enddisplaymath{\eeq\global\@ignoretrue}
\makeatother

\newtheorem{thm}{Theorem}

\newtheorem{lem}[thm]{Lemma}
\newtheorem{prop}[thm]{Proposition}
\theoremstyle{remark}
\newtheorem*{rem}{Remark}
\newtheorem{rems}{Remark}[thm]

\theoremstyle{definition}

\numberwithin{equation}{section}
\numberwithin{thm}{section}

\DeclareMathOperator{\even}{even}
\DeclareMathOperator{\odd}{odd}
\DeclareMathOperator{\alt}{alt}
\DeclareMathOperator{\fOE}{OE}
\DeclareMathOperator{\fUE}{UE}
\DeclareMathOperator{\fSE}{SE}
\DeclareMathOperator{\GOE}{GOE}
\DeclareMathOperator{\GUE}{GUE}
\DeclareMathOperator{\GSE}{GSE}
\DeclareMathOperator{\COE}{COE}
\DeclareMathOperator{\CUE}{CUE}
\DeclareMathOperator{\CSE}{CSE}
\DeclareMathOperator{\Mat}{Mat}
\DeclareMathOperator{\Symm}{Symm}
\DeclareMathOperator{\Anti}{Anti}
\DeclareMathOperator{\Beta}{Beta}
\newcommand{\R}{\mathbb R}
\newcommand{\C}{\mathbb C}
\renewcommand{\H}{\mathbb H}

\newcommand{\N}{\mathbb N}
\newcommand{\F}{\mathbb F}

\begin{document}

\title{Inter-relationships between orthogonal, unitary and
symplectic matrix ensembles}

\author{P.J.~Forrester$^\dagger$ and Eric M. Rains$^*$}
\date{}
\maketitle

\begin{center}
 $^{\dagger}$Department of Mathematics and
Statistics, University of Melbourne, \\
 Parkville, Victoria 3052, Australia \\ p.forrester@ms.unimelb.edu.au
\end{center}
\begin{center}
$^*$
AT\&T Research, Florham Park, New Jersey 07932\\
rains@research.att.com
\end{center}

\begin{abstract}
We consider the following problem:  When do alternate eigenvalues taken
from a matrix ensemble themselves form a matrix ensemble?  More precisely,
we classify all weight functions for which alternate eigenvalues from
the corresponding orthogonal ensemble form a symplectic ensemble,
and similarly classify those weights for which alternate eigenvalues from
a union of two orthogonal ensembles forms a unitary ensemble. Also
considered are the $k$-point distributions for the decimated orthogonal
ensembles.

\end{abstract}

\section{Introduction}
Given a probability measure on a space of matrices, the eigenvalue
PDF (probability density function) follows by a change of variables.
For example, consider the space of $n \times n$ real symmetric
matrices $A = [a_{j,k}]_{0\le j,k <n}$ with probability measure
proportional to
\[\label{0.1}
e^{-{\text{Tr}}(A^2)/2} (dA), \qquad (dA) := \prod_{j \le k}a_{jk}.
\]
The eigenvalues $x_0 < \cdots < x_{n-1}$ are introduced via the spectral
decomposition $A = RLR^T$ where $R$ is a real orthogonal matrix with columns
given by the eigenvectors of $A$ and $L= {\rm diag}(x_0,\dots,x_{n-1})$.
Since $e^{-\text{Tr}(A^2)/2} = e^{-\sum_{j=0}^{n-1} x_j^2/2}$ the change of
variables is immediate for the weight function; however the change of
variables in $(dA)$ cannot be carried out with such expedience.

The essential point of the latter task is to compute the Jacobian for
the change of variables from the independent elements of $A$ 
to the eigenvalues and the
independent variables associated with the eigenvectors.
Also, because only the eigenvalue PDF is being computed, one must
integrate out the eigenvector dependence. In fact the dependence in the
Jacobian on the eigenvalues separates from the dependence on the
eigenvectors, so the task of performing the integration does not become
an issue. Explicitly, one finds (see e.g.~\cite{Mu82})
$$
(dA) = \Delta(x) \prod_{j=0}^{n-1} dx_j (R^T dR)
$$
where $\Delta(x) := \Delta(x_0,\dots,x_{n-1}) := \prod_{0 \le j < k < n}
(x_k - x_j)$, and thus the eigenvalue PDF corresponding to (\ref{0.1})
is proportional to 
\[\label{0.2}
\prod_{j=0}^{n-1} g(x_j) |\Delta(x)|
\]
with $g(x) = e^{-x^2/2}$ (taking the absolute value of $\Delta(x)$ allows the
ordering restriction on the eigenvalues to be dropped).

More generally, the above working shows that a space of $n \times n$ real
symmetric matrices with probability measure proportional to 
\[ \label{0.3}
\exp \Big ( \sum_{j=1}^\infty \alpha_j {\rm Tr} (A^j) \Big ) (dA)
\]
will have eigenvalue PDF (\ref{0.2}) with $g(x) = \exp (\sum_{j=1}^\infty
\alpha_j x^j)$. Because (\ref{0.3}) is unchanged by similarity
transformations $A \mapsto RAR^T$ with $R$ real orthogonal, for general
$g$ (\ref{0.2}) is said to be the eigenvalue PDF of an orthogonal
ensemble, and denoted $\fOE_n(g)$.

Real symmetric matrices are Hermitian matrices with all elements constrained
to be real. If one considers Hermitian matrices without this constraint, so
the off diagonal elements can now be complex, the eigenvalue PDF
corresponding to the ensemble (\ref{0.3}) is again given by (\ref{0.2})
but with $|\Delta(x)|$ replaced by $(\Delta(x))^2$. Because (\ref{0.3})
is then unchanged by $A \mapsto UAU^\dagger$ for $U$ unitary, (\ref{0.2})
so modified is referred to as a unitary ensemble and denoted
$\fUE_n(g)$. The third and final possibility \cite{Dy62} is to consider
$n \times n$ Hermitian matrices in which each element is itself a $2 \times
2$ matrix of the form
\[ \label{0.4}
\left [ \begin{array}{cc} z & w \\ - \bar{w} & \bar{z} \end{array}
\right ].
\]
This class of $2 \times 2$ matrices form the real quaternion number field
$\H$.
The spectrum of such matrices, regarded as $2n \times 2n$ matrices with
complex entries, is doubly degenerate. The ensemble of matrices
(\ref{0.3}) is now invariant under the transformations $A \mapsto
BAB^\dagger$ for $B$ symplectic unitary, and so referred to as a
symplectic ensemble. The eigenvalue PDF of the distinct eigenvalues is
given by (\ref{0.2}) with $|\Delta(x)|$ replaced by $(\Delta(x))^4$, and
this is denoted $\fSE_n(g)$.

The matrix ensembles corresponding to the eigenvalue PDFs
$$
\fOE_n(e^{-x^2/2}), \quad \fUE_n(e^{-x^2}), \quad
\fSE_n(e^{-x^2})
$$
are given the special labels $\GOE_n$, $\GUE_n$ and $\GSE_n$ respectively
(the G standing for Gaussian). As seen from (\ref{0.1}) they can be realized
by an appropriate Gaussian weight function in the probability space.
Because for $A$ real symmetric
$$
e^{-{\rm Tr}(A^2)/2} = \prod_{j=0}^{n-1} e^{-a_{jj}^2} \prod_{j<k}^{n-1}
e^{-a_{jk}^2},
$$
and similarly for $A$ Hermitian with complex or real quaternion
elements, independent elements of the Gaussian ensemble are
independently distributed Gaussian random variables.

There are also a number of other
known random matrix ensembles with this latter property, and which 
have eigenvalue PDF of the form
$\fOE_n(g)$, $\fUE_n(g)$ or $\fSE_n(g)$ for some $g$. Seven such ensembles
result by taking the Hermitian part of the matrix Lie algebras related to
Cartan's ten families of infinite symmetric spaces \cite{Zi97}. We
specify five of these:
\begin{align}
\Mat(p,q;\R) & \quad\text{$p\times q$ matrices over $\R$\ \ $(p\ge q)$}\\
\Mat(p,q;\C) & \quad\text{$p\times q$ matrices over $\C$\ \ $(p\ge q)$}\\
\Mat(p,q;\H) & \quad\text{$p\times q$ matrices over $\H$\ \ $(p\ge q)$}\\
\Symm(n;\C) & \quad\text{$n\times n$ symmetric complex matrices}\\
\Anti(n;\C) & \quad\text{$n\times n$ antisymmetric complex matrices}.
\end{align}
The quantities of interest are the square of the non-zero singular values,
or equivalently the eigenvalues of $A^\dagger A$ 
for $A$ a member of the ensemble,
in each case. The first
two of these ensembles were studied long ago in mathematical statistics
\cite{Wi28,Go63}; these two together with the third have occured in
recent physical applications (see \cite{AZ97} and
references therein), while the final two (in a different
guise) have also arisen in a physical context \cite{AZ97}. The distribution
of the eigenvalues of $A^\dagger A$ can be computed in a number of ways; one
approach is to make use of the correspondence \cite{Zi97} to a symmetric
space (of types  $BDI$, $AIII$, $CII$, $CI$ and $DIII$ respectively), which
allows the tables in \cite{He62} to be utilized. Abusing notation, we have
\begin{align}\label{a.1}
\Mat(p,q;\R)   &= \fOE_q(x^{(p-q-1)/2} e^{-x/2})\nonumber \\
\Mat(p,q;\C)   &= \fUE_q(x^{p-q} e^{-x})\nonumber \\
\Mat(p,q;\H)   &= \fSE_q(x^{2(p-q)+1} e^{-x})\nonumber \\
\Symm(n;\C)    &= \fOE_n(e^{-x/2})\nonumber \\
\Anti(2n;\C)   &= \fSE_n(e^{-x})\nonumber \\
\Anti(2n+1;\C) &= \fSE_n(x^2 e^{-x})
\end{align}
Up to the scale of $x$, all the above weight functions are of the 
Laguerre form $x^\alpha e^{-x}$ and so by definition are examples of
Laguerre matrix ensembles.

Another class of matrix ensembles in which the entries of the underlying
matrices are independently distributed Gaussian random variables are known 
in mathematical statistics \cite{Mu82}. With $a \in \Mat(p_1,q;\F)$
(where $\F = \R, \C$ or $\H$), $b \in \Mat(p_2,q,\F)$, and
$A = a^\dagger a$, $B = b^\dagger b$, these distributions are described by
$$
\Beta(p_1,p_2,q;\F) \quad q \times q \: \:
\text{matrices} \: \: A(A+B)^{-1}.
$$
They have corresponding eigenvalue PDF (abusing notation as in (\ref{a.1}))
\begin{align}
\Beta(p_1,p_2,q;\R) &= \fOE_q(x^{(p_1-q-1)/2} (1-x)^{(p_2-q-1)/2}) \nonumber \\
\Beta(p_1,p_2,q;\C) &= \fUE_q(x^{p_1-q}(1-x)^{p_2-q}) \nonumber \\
\Beta(p_1,p_2,q;\H) &= \fSE_q(x^{2(p_1-q)+1} (1-x)^{2(p_2-q)+1})
\end{align}
where $0<x<1$, and thus involve weight functions of the Jacobi type.

The above revision demonstrates that it is possible to realize, in terms
of matrices with entries which are independently distributed Gaussian random
variables, the distributions $\fOE_n(g)$, $\fUE_n(g)$ and $\fSE_n(g)$ for
$g$ one of the forms
\[\label{clw}
e^{-x^2}, \quad x^{\alpha} e^{-x}, \quad x^a(1-x)^b.
\]
These same weight functions occur in the theory of orthogonal polynomials
\cite{Sz75} --- they are associated with the three families of classical
orthogonal polynomials Hermite, Laguerre and Jacobi respectively, and are
themselves referred to as classical weight functions. The classical
polynomials share many special properties not enjoyed by orthogonal
polynomials associated with other weight functions. In the present study
of matrix ensembles, we will see that the distributions $\fOE_n(g)$,
$\fUE_n(g)$ and $\fSE_n(g)$ also have special features for $g$ a classical
weight function (\ref{clw}).

Our interest is in the properties of alternate eigenvalues in matrix
ensembles. In particular we seek to determine the weights $g$ for which
alternate eigenvalues taken from a random union of two orthogonal
ensembles form a unitary ensemble. Similarly we seek the weights $g$
for which alternate eigenvalues from an orthogonal ensemble form a
symplectic ensemble. The motivation for this study comes from recent
work of Baik and Rains \cite{BR99a}. Consider the distribution
$\fOE_n(e^{-x})$, $n$ even, and order the eigenvalues $x_0<x_1<\cdots <x_{n-1}$.
In  \cite{BR99a} it was proved that after integrating out every second
eigenvalue $x_{n-1}, x_{n-3}, \dots$ etc.~the remaining eigenvalues have
the distribution $\fSE_{n/2}$. 
The proof of
Baik and Rains is particular to the $a=0$ case of the Laguerre ensemble.
However other considerations lead these authors \cite{BR99b} to conjecture
that in an appropriate scaled limit the distribution of the largest
eigenvalue in the GSE corresponds to that of the second largest eigenvalue
in the GOE. From this it is remarked that presumably the joint distribution
of every second eigenvalue in the GOE coincides with the joint distribution
of all the eigenvalues in the GSE, with an appropriate number of eigenvalues.

Baik and Rains \cite{BR99b} were also led to consider two $\GOE_n$ spectra,
superimposing them at random, and integrating out every second eigenvalue
of the resulting sequence. Results were presented which suggest 
that in the scaled $n \to \infty$ limit at the soft edge the distribution
becomes that of $\GUE_\infty$, appropriately scaled.

Such inter-relationships between ensembles first occured in the work of
Dyson \cite{Dy62c} on the circular ensembles of random unitary matrices.
This ensemble has eigenvalue PDF proportional to
\[\label{0.8}
\prod_{0 \le j < k < n} |e^{i\theta_k} - e^{i\theta_j}|^\beta,
\quad 0 \le \theta_j < 2 \pi
\]
for $\beta = 1,2$ and 4 ($\COE_n$, $\CUE_n$ and $\CSE_n$) respectively.
Dyson conjectured that
\[\label{0.9}
\text{alt}(\COE_n \cup \COE_n) = \CUE_n
\]
which means that if two spectra from the
$\COE_n$ distribution are superimposed at random with
every second eigenvalue integrated out, the $\CUE_n$ distribution
results. This was subsequently proved by Gunson \cite{Gu62}. Also,
Mehta and Dyson \cite{MD63} proved that integrating out every second
eigenvalue from the distribution $\COE_n$ with $n$ even gives the
distribution $\CSE_{n/2}$, or symbolically
\[\label{0.10}
\text{alt}(\COE_n) = \CSE_{n/2}.
\]

The circular ensembles can be analyzed in the course of the present study of
ensembles with real valued eigenvalues by making the stereographic
projection
$$
e^{i \theta_j} = \frac{1 - i x_j}{1 + i x_j}.
$$
The PDF (\ref{0.8}) then maps to
$$
\prod_{j=0}^{n-1} \frac{1}{(1 + x_j^2)^{\beta(n-1)/2 + 1}}
\prod_{0 \le j < k < n} |x_k - x_j|^\beta
$$
which is of the general type under consideration. Here the weight function is
of the form
\[\label{clw1}
\frac{1}{(1 + x^2)^\alpha}, \quad \alpha > 1.
\]
This only has a finite number of well defined moments and thus in this
respect differs from the classical weight functions (\ref{clw}).
On the other hand the corresponding orthogonal polynomials are
$\{P_n^{(-\alpha,-\alpha)}(ix)\}_{n< \alpha - 1/2}$ \cite{Ro29},
with $P_n^{(\alpha,\beta)}$ denoting the Jacobi polynomial,
thus implying (\ref{clw1}) can be viewed as a fourth classical weight function.

\section{Pseudo-ensembles}\label{sec:pseudo}

We begin with the orthogonal ensemble eigenvalue PDF (\ref{0.2}), taking
away the modulus sign, replacing $n$ by $l$ (to avoid overuse of
the former) and rewriting the product as a determinant using
the Vandermonde formula to obtain
\[
\Delta(x) \prod_i g(x_i) 
=
\det(x_i^j)_{0\le i,j<l} \prod_i g(x_i)
=
\det(g(x_i) x_i^j)_{0\le i,j<l}.
\]
In particular, we note that each row corresponds to a variable,
while each column corresponds to a function.
Given a collection of $n$ functions $F_i:\R\to \R$,
we thus define the associated ``orthogonal pseudo-ensemble'' by the
following ``density'':
\[
\det(F_j(x_i))_{0\le i,j<l}.
\]
Thus any orthogonal ensemble is also an orthogonal pseudo-ensemble,
but certainly not vice versa.  Indeed, one has:

\begin{thm}\label{thm:pOE=OE}
Fix an integer $l>0$, and let $G:\R\to \R$ be a function supported
on at least $n$ points.  Then for a collection of $n$ functions
$F_0$, $F_1$,\dots $F_{l-1}$, we have
\[
\det(F_j(x_i))_{0\le i,j<l}
\propto
\prod_i G(x_i) \Delta(x)
\label{eq:pOE=OE}
\]
if and only if there exist $l$ linearly independent polynomials $p_i$ of
degree at most $l-1$ such that $F_i(x)=p_i(x)G(x)$ for all $i$ and
$x$.
\end{thm}

\begin{proof}
The ``if'' portion is easy enough:
\[
\det(G(x_i) p_j(x_i))_{0\le i,j<l}
=
\prod_i G(x_i) \det(p_j(x_i))_{0\le i,j<l}
\propto
\prod_i G(x_i) \Delta(x),
\]
since the polynomials are assumed linearly independent.

Now, suppose \eqref{eq:pOE=OE} holds.  It will turn out to be
convenient to restate the equation in terms of exterior products.
Define a vector-valued function $V_F(x)$ by
\[
V_F(x)_i = F_i(x).
\]
Then we can write
\[
\det(F_j(x_i))_{0\le i,j<l}
=
\langle V_F(x_0)
,
(\bigwedge_{1\le i<l} V_F(x_i))\rangle,
\]
where $\langle,\rangle$ stands for the standard duality between $1$-forms and
$l-1$-forms.  Consider this as a function of $x_0$ as the other variables
range over the support of $G$; we have:
\[
V_F(x)
\cdot
(\bigwedge_{1\le i<l} V_F(x_i))
\propto
G(x) \prod_{1\le i<l} (x-x_i).
\]
Now, since $G$ has at least $l$ elements in its support, these functions
span an $l$-dimensional space (this follows, for instance, from Lagrange's
interpolation formula).  On the other hand, the functions must clearly be
linear combinations of the $F_i$.  Since there only $l$ functions $F_i$, it
follows that we can write the $F_i$ as linear combinations of the functions
$G(x) x^i$, $0\le i<l$.  But this is precisely what we wanted to prove.
\end{proof}

Similarly, the density function of a symplectic ensemble can also be
written as a determinant, namely
\[
\Delta(x)^4 \prod_i g(x_i)^2
=
\det(g(x_i) x_i^j, j g(x_i) x_i^{j-1})_{0\le i<l, 0\le j<2l};
\]
this follows by differentiating the Vandermonde determinant.  When
$\log(g)$ is differentiable, we can perform column transformations to put
this determinant in the form
\[
\det( F_j(x_i), F'_j(x_i) )_{0\le i<l, 0\le j<2l};
\label{eq:pSE}
\]
simply take $F_j(x) = g(x) x^j$, and observe that
\[
F'_j(x) - {g'(x)\over g(x)} F_j(x) = j g(x) x^{j-1}.
\]
In fact, we can often define \eqref{eq:pSE} even when the functions
$F_j$ are not differentiable, by expressing it in terms of the
$2$-form-valued function
\[
V^{(2)}_F(x)=\lim_{y\to x} {1\over (x-y)}(V_F(x)\wedge V_F(y)).
\]
For $F_j(x)=g(x) x^j$, we find that this is defined wherever $g$ is
continuous.

\begin{thm}\label{thm:pSE=SE}
Fix an integer $l>1$, let $O$ be a nonempty open subset of $\R$, and let
$G:O\to \R$ be a continuous function supported on $O$.  Then for a
collection of continuous functions $F_j:O\to \R$ such that \eqref{eq:pSE}
is well-defined,
\[
\det( F_j(x_i), F'_j(x_i) )_{0\le i<l, 0\le j<2l}
=
\Delta(x)^4 \prod_i G(x_i)^2
\]
on $O^{2l}$ if and only if there exist linearly independent polynomials
$p_j$ of degree at most $2l-1$ with $F_j(x)=G(x) p_j(x)$.
\end{thm}

\begin{proof}
Again the ``if'' case is straightforward.  In the other direction, we can
clearly divide each $F_j$ by $G$, and thus may assume WLOG that $G=1$ on
$O$.

We first consider the case $l=2$, for which
\[
\det( F_j(x)\ F'_j(x)\ F_j(y)\ F'_j(y))_{0\le j<4}
=
\langle V^{(2)}_F(x) , V^{(2)}_F(y)\rangle
=
(x-y)^4.
\]
As $y$ varies over $O$, this spans a 5-dimensional function space; it
follows that as $y$ varies, $V^{(2)}_F(y)$ spans a 5-dimensional space (the
dimension must be either 5 or 6; 6 clearly leads to a contradiction).  In
other words, there must be a linear dependence between the coefficients of
$V^{(2)}_F(y)$.  By replacing the $F_i$ with an orthogonal linear
combination, we find that this dependence is WLOG of the form
\[
V^{(2)}_F(y)_{01} = C V^{(2)}_F(y)_{23},
\]
for some constant $C$.  Now, if $C$ were 0, then we would have
\[
F_0(x) F'_1(x) = F_1(x) F'_0(x).
\label{eq:pSE=SE.1}
\]
Now, let $I\subset O$ be an open interval in $O$.  If either $F_0$ or $F_1$
were identically 0 on $I$, our determinant would be identically 0 on $I^2$
(contradiction); it follows that we may choose $I$ so that both $F_0$ and
$F_1$ are nonzero.  Then we can divide both sides of \eqref{eq:pSE=SE.1}
by $F_0(x)F_1(x)$ and integrate; we find that $F_0\propto F_1$ on $I$.  But
this again makes the determinant 0.  We conclude that the linear dependence
satisfied by $V^{(2)}_F$ must take the form
\[
V^{(2)}_F(y)_{01} = C V^{(2)}_F(y)_{23}
\]
with $C\ne 0$.

In particular, we find that the $2$-form $V'$ orthogonal to $V^{(2)}_F(y)$
is not itself in the span of $V^{(2)}_F(y)$.  In particular, any
$2$-form can be written as a linear combination of $V'$ and some of the
$V^{(2)}_F(y)$.  Taking the inner product with $V^{(2)}_F(x)$, we conclude
that for $0\le i<j\le 3$, we have
\[
V^{(2)}_F(x)_{ij} = p_{ij}(x)
\]
for some polynomial $p$ of degree at most 4.  Now, since
$V_F(x)\wedge V^{(2)}_F(x)=0$, we find:
\[
F_0(x) p_{12}(x)- F_1(x) p_{02}(x)+ F_2(x) p_{01}(x)=0
\]
for all $i,j,k$.  Similarly, since
\[
{d\over dx} V^{(2)}_F(x) = V_F(x) \wedge V''_F(x)
\],
we have
\[
F_0(x) p'_{12}(x)- F_1(x) p'_{02}(x)+ F_2(x) p'_{01}(x)=0
\]
Now, since $V^{(2)}_F(x)_{ki}$ is linearly
independent of $V^{(2)}_F(x)_{ij}$, we can solve these two
equations for $F_1$ and $F_2$ as rational multiples of $F_0$,
subsitute into the equation $V^{(2)}_F(x)_{12}=p_{12}(x)$,
then solve for $F_0$.  We find:
\begin{align}
F_0 &= {
p_{01}p'_{02}-p_{02}p'_{01}
\over
\sqrt{D}}\\
F_1 &= {
p_{01}p'_{12}-p_{12}p'_{01}
\over
\sqrt{D}}\\
F_2 &= {
p_{02}p'_{12}-p_{12}p'_{02}
\over
\sqrt{D}}
\end{align}
where
\[
D = \det\pmatrix
p_{01}&p_{12}&p_{20}\\
p'_{01}&p'_{12}&p'_{20}\\
p''_{01}&p''_{12}&p''_{20}\endpmatrix
\]
We observe that each numerator has degree at most 6, as does the polynomial
$D$.  In particular, if we exclude any given $F$, we can express the
squares of the other $F$ as rational functions with common denominator of
degree at most 6.  It follows that the functions $F^2$ have at most 8 poles
between them, and thus that we can write
\begin{align}
F_0(x) &= p_0(x) p(x)^{-1/2}\\
F_1(x) &= p_1(x) p(x)^{-1/2}\\
F_2(x) &= p_2(x) p(x)^{-1/2}\\
F_3(x) &= p_3(x) p(x)^{-1/2},
\end{align}
where $p_0$, $p_1$, $p_2$, and $p_3$ are polynomials of degree at most 7
and $p$ is a polynomial of degree at most 8.

We now need to show that, in fact, each $F_i$ is a polynomial of degree
at most 3.  By the usual factorization, we find:
\[
\det(p_j(x)\ p'_j(x)\ p_j(y)\ p'_j(y))_{0\le j<4}
\propto
p(x) p(y) (x-y)^4,
\label{eq:pSE=SE.2}
\]
valid on $\R$.  Without loss of generality, we may assume that the constant
of proportionality is 1, and that $p(0)=1$.  Dividing both sides by
$(x-y)^4$ and taking the limit as $x,y\to 0$, we find:
\[
\det\pmatrix
p_0(0)&p_1(0)&p_2(0)&p_3(0)\\
p'_0(0)&p'_1(0)&p'_2(0)&p'_3(0)\\
p''_0(0)&p''_1(0)&p''_2(0)&p''_3(0)\\
p'''_0(0)&p'''_1(0)&p'''_2(0)&p'''_3(0)
\endpmatrix
=
1
\]
Applying a suitable linear transformation to the polynomials $p_i$, we
have, without loss of generality,
\begin{align}
p_0(x) &= 1+p_{04}x^4+p_{05}x^5+p_{06}x^6+p_{07}x^7  \\
p_1(x) &= x+p_{14}x^4+p_{15}x^5+p_{16}x^6+p_{17}x^7  \\
p_2(x) &= x^2+p_{24}x^4+p_{25}x^5+p_{26}x^6+p_{27}x^7\\
p_3(x) &= x^3+p_{34}x^4+p_{35}x^5+p_{36}x^6+p_{37}x^7
\end{align}
We can then solve for $p(x)$ by taking $y=0$ above; we find
\[
p(x) = x^{-4} (p_2(x) p'_3(x)-p_3(x) p'_2(x)).
\]
At this point, we can compare coefficients on both sides of
\eqref{eq:pSE=SE.2}, obtaining a number of polynomial equations relating
the coefficients $p_{ij}$, $0\le i\le 3$, $4\le j\le 7$.
The resulting ideal can be verified
(using {\tt magma}, for instance) to contain the polynomials
$(p_{25}+p_{34}p_{35}-p_{36})^2$, $(p_{26}+p_{34}p_{36}-p_{37})^2$, and
$(p_{27}+p_{34}p_{37})^2$; passing to the radical, we can then solve
for $p_{ij}$, $0\le i\le 2$, $4\le j\le 7$.
Substituting in, we find that $p(x)$ is now a square, and that
each $p_i(x)$ is a multiple of $\sqrt{p(x)}$.  In other words, each
$F_i(x)$ is a polynomial of degree at most 3, and we are done with the case
$l=2$.

It remains only to show that we can reduce the cases $l>2$ to cases of
lower dimension.  Choose a particular element $x_0\in O$.  By replacing the
$F_i$ with appropriate linear transformations, we may assume
\[
V^{(2)}_F(x_0) = C e_1\wedge e_2,
\]
for some nonzero constant $C$.  In particular, we find that
\[
\det(F_j(x_i))_{0\le i<l,\ 0\le j\le 2l}
=
C \det(F_j(x_i))_{1\le i<l,\ 2\le j\le 2l}
\propto
G(x_0)
\Delta(x_1,x_2,\dots x_{l-1})^4 \prod_{1\le i<l} x^4_i G(x_i)^2.
\]
By induction, it follows that for $2\le i<l$, there
exist polynomials $p_i(x)$ of degree at most $2l-1$ such that
$F_i(x)=(x-x_0)^2 G(x) p_i(x)$ on $O$.  Undoing our linear transformations,
we find that for every polynomial $p(x)$ of degree at most $2l-1$
vanishing to second order at $x_0$, we can write $G(x)p(x)$ as
a linear combination of the $F_i(x)$.  But this was independent of
our choice of $x_0$.  In particular, taking $x'_0$ to be any other element
of $O$, we have
\[
1 = {3 (x_0-x'_0)(x-x_0)^2 +2 (x-x_0)^3+3 (x_0-x'_1)(x-x'_0)^2-2 (x-x'_0)^3\over (x_0-x'_0)^3}
\]
and
\[
(x-x_0) = {-2(x_0-x'_0)(x-x_0)^2
           - (x-x_0)^3
           - (x_0-x'_0) (x-x'_0)^2
           + (x-x'_0)^3\over (x_0-x'_0)^2}.
\]
It follows that for any polynomial $p(x)$ of degree at most $2l-1$,
$G(x)p(x)$ is a linear combination of the $F_i(x)$.  By dimensionality,
it follows that each $F_i(x)$ is itself of the form $G(x)p(x)$, and
we are done.
\end{proof}

\section{Linear fractional transformations}

It will be convenient in the sequel to determine how matrix ensembles
behave under a linear fractional change of variables.  To be precise, let
$f$ be a weight function, and consider what the density of one of its
associated matrix ensemble is in terms of the variables $y_i$ defined by
$x_i = (\alpha y_i +\beta)/(\gamma y_i+\delta)$.  Clearly, we
need only determine how $\Delta(x)$ and $\prod_i dx_i$ transform.

We readily compute:
\[
dx = {\alpha\delta-\beta\gamma\over (\gamma y+\delta)^2},
\]
thus answering that question.  As for $\Delta$:

\begin{lem}
Let $y_0$, $y_1$, \dots $y_{l-1}$ be a collection of
$l$ real numbers.  Then for any $\alpha$, $\beta$, $\gamma$ and
$\delta$ such that $\gamma y_i+\delta$ is never 0,
\[
\Delta\left({\alpha y_i+\beta\over \gamma y_i+\delta}\right)
=
(\alpha\delta-\beta\gamma)^{l(l-1)/2}
\prod_i (\gamma y_i+\delta)^{1-l}
\Delta(y_i).
\]
\end{lem}

\begin{proof}
For each $i<j$, we have
\[
{\alpha y_j+\beta\over \gamma y_j+\delta}-{\alpha y_i+\beta\over \gamma
y_i+\delta}
=
{\alpha\delta-\beta\gamma\over (\gamma y_i+\delta)(\gamma y_j+\delta)}
(y_j-y_i).
\]
Multiplying over $i<j$, we are done.
\end{proof}

We thus obtain the following transformation rules:

\begin{thm}
Let $f$ be any weight function.  Under the change of variables
$x_i=(\alpha y_i+\beta)/(\gamma y_i+\delta)$, we have:
\begin{align}
\fOE_l(f(x)) &\to \fOE_l(
|\alpha\delta-\beta\gamma|^{(l+1)/2}
(\gamma y+\delta)^{-1-l}
\tilde{f}(y))
\\
\fUE_l(f(x)) &\to \fUE_l(
|\alpha\delta-\beta\gamma|^{l}
(\gamma y+\delta)^{-2l}
\tilde{f}(y))
\\
\fSE_l(f(x)) &\to \fSE_l(
|\alpha\delta-\beta\gamma|^{2l-1}
(\gamma y+\delta)^{2-4l}
\tilde{f}(y)),
\end{align}
where $\tilde{f}(y) = f((\alpha y+\beta)/(\gamma y+\delta))$, the
normalization constants are the same on both sides, and for
$\fOE_{2l}$, $\gamma y+\delta$ must be positive over the support of
$\tilde{f}$.
\end{thm}

\begin{proof}
When $\alpha\delta-\beta\gamma<0$, the LFT reverses the order of
integration, thus justifying the extra factor of $(-1)^l$ introduced for
$\fUE_l$ and $\fSE_l$.  For $\fOE_l$, there is a more subtle difficulty,
namely that the relative order of the eigenvalues is significant, and can
change.  If we simply reverse the order, this is not a problem (the total
effect is $(-1)^{l(l+1)/2}$, thus cancelling out the sign of
$\alpha\delta-\beta\gamma$).  So we can restrict to the case
$\alpha\delta-\beta\gamma>0$.  The effect of the LFT is then to cyclically
shift the ordering, taking the eigenvalues with $x>\alpha/\gamma$ and
making them smallest.  If there are $k$ such eigenvalues, the
sign of the Vandermonde matrix is changed by $(-1)^{k(l-1)}$; thus
if $l$ is odd, there is no problem.  On the other hand, if $l$ is odd,
we have a problem unless the eigenvalues are restricted to only
one side of $\alpha/\gamma$, or equivalently that $\gamma y+\delta$ has
constant sign over the support of $\tilde(f)$.  Since
\[
{\alpha y+\beta\over \gamma y+\delta}
=
{-\alpha y-\beta\over -\gamma y-\delta}
\]
we may take this sign to be positive.
\end{proof}

\begin{rem}
For algebraic purposes, we can often ignore the constraint $\gamma
y+\delta>0$, since the transform still has the correct form to be
a matrix ensemble density, despite not being nonnegative.
\end{rem}

The upshot of this is that we can use this freedom to send a suitably
chosen point to $\infty$, thus simplifying our analysis below.

\section{The main results}\label{s3}

For a matrix ensemble $M$, we define $\even(M)$ to be the ensemble obtained
by taking the $2$nd largest, $4$th largest, etc. eigenvalues of $M$, and
similarly for $\odd(M)$.

When considering $\even(M)$ or $\odd(M)$ for $M=\fOE_n\cup \fOE_n$ or
$M=\fOE_n\cup \fOE_{n+1}$, the following lemma is crucial:

\begin{lem}
For any integer $n>0$,
\[
\sum_{\substack{S\subset \{0,1,\dots 2n-1\}\\|S|=n}}
\Delta(x_S)\Delta(x_{\{0,1,\dots 2n-1\}-S})
=
2^n
\Delta(x_{\{0,2,\dots 2n-2\}})
\Delta(x_{\{1,3,\dots 2n-1\}})
\]
and
\[
\sum_{\substack{S\subset \{0,1,\dots 2n\}\\|S|=n+1}}
\Delta(x_S)\Delta(x_{\{0,1,\dots 2n\}-S})
=
2^n
\Delta(x_{\{0,2,\dots 2n\}})
\Delta(x_{\{1,3,\dots 2n-1\}})
\]
\end{lem}

\begin{proof}
Consider what happens when we exchange $x_i$ and $x_{i+2}$ in a
term of either equation.  If $i,i+2\in S$ or $i,i+2\not\in S$,
then
\[
\Delta(x_S)\Delta(x_{\{0,1,\dots l-1\}-S})
\to
-\Delta(x_S)\Delta(x_{\{0,1,\dots l-1\}-S}),
\]
since $\Delta$ is alternating.  Otherwise, we see that every factor
$x_j-x_k$ with $j>k$ is taken to another such factor, {\it except}
for the factor $x_{i+1}-x_i$ or $x_{i+2}-x_{i+1}$, whichever is
present.  So each term in our sum is taken to the negative of a
term from our sum; it follows that the sum is alternating under
parity-preserving permutations.  It follows that it must be a multiple
of
\[
\Delta(x_{\{0,2,\dots 2\lceil l/2\rceil-2\}})
\Delta(x_{\{1,3,\dots 2\lfloor l/2\rfloor-1\}}).
\]
By degree considerations, it remains only to verify the constant,
which we can do by considering the coefficient of largest degree in 
$x_0$, and applying induction.
\end{proof}

\begin{rem}
The even case of this lemma is implicit in \cite{Gu62}, where it was
used to analyze $\even(\fOE_n\cup \fOE_n)$ with respect to the
weight function $1$ on the unit circle.
\end{rem}
From the lemma, it follows that the density of $\fOE_n(f)\cup \fOE_n(f)$,
expressed in terms of ordered variables, is proportional to
\[
\prod_{0\le i\le 2n} f(x_i)
\Delta(x_{\{0,2,\dots 2n-2\}})
\Delta(x_{\{1,3,\dots 2n-1\}});
\]
similarly, the density of $\fOE_n(f)\cup \fOE_{n+1}(f)$ is proportional to
\[
\prod_{0\le i\le 2n} f(x_i)
\Delta(x_{\{0,2,\dots 2n\}})
\Delta(x_{\{1,3,\dots 2n-1\}}).
\]
For some weight functions $f$, if we integrate over the odd/even variables,
the resulting density is the density of a unitary ensemble; we wish to
determine precisely when that is.  We first consider the case
$\even(\fOE_2(f)\cup \fOE_3(f))$.

\begin{thm}\label{thm:OEOE.1}
Let $f:\R\to \R$ be a function which is differentiable on a possibly
unbounded open interval $I\subset \R$ and 0 elsewhere.  Suppose
$\even(\fOE_2(f)\cup \fOE_3(f))=\fUE_2(g)$ for some function $g$.  Then up
to a linear transformation of variables, $f$ must have one of the following
forms.  On the interval $(0,1)$:
\begin{align}
f(x) &\propto x^\alpha (1-x)^\beta (1-rx)^{-4-\alpha-\beta},\quad
\alpha,\beta>-1,\ r<1\\
f(x) &\propto x^{-4-\alpha} e^{-1/x} (1-x)^{\alpha},\quad
\alpha>-1.
\end{align}
On the interval $(0,\infty)$:
\begin{align}
f(x) &\propto x^\alpha (1-rx)^\beta,\quad
\alpha>-1,\ \alpha+\beta<-3,\ r<0\\
f(x) &\propto x^{-4-\alpha} e^{-1/x},\quad \alpha>-1,\\
f(x) &\propto x^{\alpha} e^{-x},\quad \alpha>-1.
\end{align}
Finally, on the entire real line:
\begin{align}
f(x) &\propto (1+x^2)^\alpha,\quad \alpha<-3/2\\
f(x) &\propto e^{-x^2/2}.
\end{align}
\end{thm}

\begin{proof}
We need to integrate this over the variables $x_{2i}$, and thus need
to evaluate the determinant
\[
\det(
\int_{[x_{2i-1},x_{2i+1}]} f(x) x^j\, dx
)_{0\le i,j\le 2},
\]
where we take $x_{-1}=a$ to be the left endpoint of $I$, and $x_{5}=b$
to be the right endpoint of $I$.  In particular, we need to determine
when there exists a function $g(x)$ with
\[
\det(
\int_{[x_{2i-1},x_{2i+1}]} f(x) x^j\, dx
)_{0\le i,j\le 2}
\propto
g(x_1)g(x_3) (x_3-x_1).
\]
As in \cite{Me91}, section 10.6,
we may use row operations to transform this to:
\[
\det(
F_j(x_{2i+1})
)_{0\le i,j\le 2}),\
\]
where we define
\[
F_j(y)=\int_{[a,y]} f(x) x^j \,dx.
\]
We cannot quite apply theorem \ref{thm:pOE=OE}, however, since the last
column of our determinant is constant.  However, we clearly have
$F_0(b)>0$, so we can eliminate that column, obtaining
\[
F_0(b)
\det(
F_j(x_{2i+1})-{F_j(b)\over F_0(b)} F_0(x_{2i+1})
)_{0\le i,j\le 1}).
\]
This, then, satisfies the hypotheses of theorem \ref{thm:pOE=OE}; there
thus exist linear polynomials $p_j$ such that
\[
p_2(x) (F_1(x)-C_1 F_0(x)) = p_1(x) (F_2(x)-C_j F_0(x)),
\label{eq:OEOE.1.1}
\]
where we have set
\[
C_i = {F_i(b)\over F_0(b)}.
\]

Differentiating twice and using the definition of $F_i$, we find:
\[
(p_2(x)(x-C_1)-p_1(x)(x^2-C_2)) f'(x)
=
(-2(x-C_1)p'_2(x)+2(x^2-C_2) p'_1(x)-p_2(x)+2x p_1(x)) f(x).
\]
We can thus solve this for $f'(x)/f(x)$; we find that $f'(x)/f(x)$ has the
form $p(x)/q(x)$ with $\deg(p)\le 2$, $\deg(q)\le 3$, and $\deg(xp+4q)\le
2$.  We observe that these conditions are, naturally, preserved by linear
fractional transformations.  In particular, by applying a suitable
linear fractional transformation, we may insist that $q$ be strictly cubic,
and that both endpoints of $I$ be finite (possibly equal).  (The
result may very well no longer be a matrix ensemble, but as we noted
above, this does not affect any algebraic conclusions.)

Now, consider how $f(x)$ and $q(x)$ must behave at $0$ and $1$
Differentiating \eqref{eq:OEOE.1.1} once and taking a limit $x\to x_{-1}$
we find, since each $F_i(x_{-1})=0$,
\[
\lim_{x\to x_{-1}}
(p_2(x) (x-C_1)+p_1(x) (C_2-x)) f(x) = 0
\]
But this is just $\lim_{x\to x_{-1}} q(x)f(x)$.  If $q(x_{-1})\ne 0$,
then we must have $\lim_{x\to x_{-1}} f(x)=0$.  Then
\[
\lim_{x\to x_{-1}} {f'(x)\over f(x)} = \infty.
\]
The only way this can happen is if $q(x_{-1})=0$ after all.  Similarly,
we have $q(x_{2n+1})=0$.

Suppose first that $a\ne b$.  Then up to LFT, we may insist that $a=0$ and
$b=1$, and thus $q(0)=q(1)=0$.  We thus have two possibilities.  The
first is that $q(x)$ has an additional zero, neither 0 nor 1.  In
this case, integrating $f'/f$ and taking into account the constraints
on $p(x)$, we obtain
\[
f(x) = x^{\alpha} (1-x)^{\beta} (1-rx)^{-4-\alpha-\beta}
\]
Now, for $\int f(x)$ not to diverge at $0$, we must have $\alpha>-1$,
and similarly $\beta>-1$.  But then $-4-\alpha-\beta<-2$; it follows
that $(1-rx)$ must be nonzero on $(0,1)$; in particular, $r<1$.
The other possibility is that $q(x)$ has a double root, WLOG at 0. 
Upon integrating $f'/f$, we obtain
\[
f(x) = x^{-4-\alpha} e^{-\beta/x} (1-x)^{\alpha},
\]
and find $\alpha>-1$, $\beta>0$. The possibilities for $(0,\infty)$ then
follow by LFT.

The other case we must consider is $I=\R$, and thus $\deg(q)=2$,
$\deg(p)=1$.  If $q$ had a simple root in $\R$, it would have two (WLOG 0 and
1), and thus $f$ would have the form $x^\alpha (1-x)^\beta$ with
$\alpha,\beta>-1$.  But then the integral for $F_2$ would diverge.
Similarly, if $q$ had a double root at 0, $f$ would have the
form $x^\alpha e^{-\beta/x}$, which would diverge on one side of $0$.
Thus either $q$ has a pair of complex roots, or $\deg(q)=0$.
In the first case, a linear transformation takes the roots to $\pm i$,
and thus
\[
f(x) = (1+x^2)^\alpha.
\]
For $F_2$ to be well-defined, we must have $\alpha<-3/2$.
The other possiblity gives $\log(f(x)) = a x^2+b x+c$; thus a linear
transformation gives
\[
f(x) = e^{-x^2/2}.
\]
\end{proof}

We can now extend this to $n\ge 2$.

\begin{thm}\label{thm:OEOE.2}
Fix an integer $n\ge 2$.  Let $f:\R\to \R$ be a function which is
differentiable on a possibly unbounded open interval $I\subset \R$ and 0
elsewhere.  Suppose $\even(\fOE_n(f)\cup \fOE_{n+1}(f))=\fUE_n(g)$ for some
function $g$.  Then up to a linear transformation of variables, $f$ and $g$
can have precisely the following forms.  On the interval $(0,1)$:
\begin{align}
f(x) &\propto x^\alpha (1-x)^\beta (1-rx)^{-n-2-\alpha-\beta},\quad
\alpha,\beta>-1,\ r<1\\
g(x) &\propto x^{2\alpha+1} (1-x)^{2\beta+1}
(1-rx)^{-2n-3-2\alpha-2\beta}\\
\noalign{\smallskip}
f(x) &\propto x^{-n-2-\alpha} e^{-1/2x} (1-x)^{\alpha},\quad
\alpha>-1\\
g(x) &\propto x^{-2n-2-2\alpha} e^{-1/x} (1-x)^{2\alpha+1}.
\end{align}
On the interval $(0,\infty)$:
\begin{align}
f(x) &\propto x^\alpha (1-rx)^\beta,\quad \alpha>-1,\ \alpha+\beta<-n-1,\ r<0\\
g(x) &\propto x^{2\alpha+1} (1-rx)^{2\beta+1}\\
\noalign{\smallskip}
f(x) &\propto x^{-n-2-\alpha} e^{-1/2x},\quad \alpha>-1\\
g(x) &\propto x^{-2n-2-2\alpha} e^{-1/x}\\
\noalign{\smallskip}
f(x) &\propto x^{\alpha} e^{-x/2},\quad \alpha>-1\\
g(x) &\propto x^{2\alpha+1} e^{-x}.
\end{align}
Finally, on the entire real line:
\begin{align}
f(x) &\propto (1+x^2)^\alpha,\quad \alpha<-(n+1)/2\\
g(x) &\propto (1+x^2)^{2\alpha+1}\\
\noalign{\smallskip}
f(x) &\propto e^{-x^2/2}\\
g(x) &\propto e^{-x^2}.
\end{align}
\end{thm}

\begin{proof}
As for $n=2$, the issue is when
\[
\det(
F_j(x_{2i+1})
)_{0\le i,j\le n})
\]
takes the form of an orthogonal ensemble.  Applying an LFT as necessary, we
may assume that $a=-\infty$.  Now differentiate with respect to $x_1$,
divide by $x_1^{n-1} f(x_1)$, and take a limit as $x_1\to -\infty$.  On the
one hand, this operation takes orthogonal ensembles to orthogonal
ensembles.  On the other hand, we can then expand along the first column,
finding that
\[
\det(
F_{j-1}(x_{2i+1})
)_{1\le i,j\le n})
\]
must take the form of an orthogonal ensemble.  By induction, we
find that $f(x)$ must satisfy the constraints valid for $n-1$.
Upon undoing the LFT, we obtain the desired ``only if'' result.

It remains to show that each of the above weight functions actually
do work.  We need only consider the following possibilities:
\begin{align}
f(x) &= x^\alpha (1-x)^\beta \label{3.44}\\
f(x) &= x^\alpha e^{-x}\\
f(x) &= (1+x^2)^\alpha\\
f(x) &= e^{-x^2/2} \label{3.47},
\end{align}
(on $(0,1)$, $(0,\infty)$, $\R$, and $\R$ respectively) since the others
are all images of these under LFTs.

For $f(x)=x^\alpha (1-x)^\beta$, observe that
\[
(\alpha+\beta+j+2) F_{j+1}(x)
-
(\alpha+j+1) F_j(x)
=
-x^j (x^{\alpha+1} (1-x)^{\beta+1});
\]
this is true for $x=0$, and both sides have the same derivative.
In particular, for each $j$, we have a polynomial $p_j(x)$ of
degree $j-1$ and a constant $C_j$ with
\[
F_j(x) = C_j F_0(x) + p_j(x) (x^{\alpha+1} (1-x)^{\beta+1}).
\]
In particular, this must be true for $x=1$, and thus $C_j=F_j(1)/F_0(1)$
as required.  We thus find that we obtain a unitary ensemble with
weight function proportional to $x^{2\alpha+1} (1-x)^{2\beta+1}$.
Similarly, for $f(x)=x^\alpha e^{-x}$, we have
\[
F_{j+1}(x)
-
(\alpha+j+1) F_j(x)
=
-x^{\alpha+j+1} e^{-x},
\]
so $g(x)\propto x^{2\alpha+1} e^{-2x}$.

For $f(x)=(1+x^2)^\alpha$, we find
\[
(2\alpha+j+2) F_{j+1}(x)
+
jF_{j-1}
=
x^j (1+x^2)^{\alpha+1}
\]
This allows us to solve for each $F_j$ except $F_0$; we obtain
$g(x)\propto (1+x^2)^{2\alpha+1}$.
Finally, for $f(x)=e^{-x^2/2}$, we have:
\[
F_{j+1}
-
j F_{j-1}
=
-x^j e^{-x^2/2},
\]
and $g(x)\propto e^{-x^2}$.
\end{proof}

\begin{rem}
We observe that in each case $g(x)\propto f(x)^2 q(x)$.
\end{rem}

For $\even(\fOE_n\cup \fOE_n)$, the calculations are analogous,
and we have:

\begin{thm}\label{thm3.4}
Fix an integer $n\ge 2$.  Let $f:\R\to \R$ be a function which is
differentiable on a possibly unbounded open interval $I\subset \R$ and 0
elsewhere.  Suppose $\even(\fOE_n(f)\cup \fOE_n(f))=\fUE_n(g)$ for some
function $g$.  Then up to an order-preserving linear transformation of
variables, $f$ and $g$ must have one of the following forms.  On 
an interval with right endpoint $0$:
\begin{align}
f(x) &\propto (-x)^\alpha (1-rx)^{-n-1-\alpha},\quad \alpha>-1,\ 1\notin rI\\
g(x) &\propto (-x)^{2\alpha+1} (1-rx)^{-2n-1-2\alpha}\\
\noalign{\smallskip}
f(x) &\propto (-x)^{-n-1} e^{1/x}\\
g(x) &\propto (-x)^{-2n} e^{2/x}.
\end{align}
On an interval of the form $(a,\infty)$, $a>-\infty$:
\begin{align}
f(x) &\propto (1-rx)^\alpha,\quad \alpha<-n,\ r<0\\
g(x) &\propto (1-rx)^{2\alpha+1}\\
\noalign{\smallskip}
f(x) &\propto e^{-x},\\
g(x) &\propto e^{-2x}
\end{align}
On the entire real line, no possibilities exist.
\end{thm}

\begin{rem}
The relation between $g$ and $f$ is here slightly modified, by removing
the factor of $q$ corresponding to the left endpoint; similarly, for
$\odd(\fOE_n\cup \fOE_n)$, we remove the factor corresponding to the
right endpoint, and for $\odd(\fOE_{n-1}\cup\fOE_n)$, we remove
both factors.
\end{rem}

For $\odd(\fOE_n\cup \fOE_n)$, we need simply reverse the ordering.
For $\odd(\fOE_n\cup \fOE_{n+1})$, we have:

\begin{thm}\label{thm3.5}
Fix an integer $n\ge 2$.  Let $f:\R\to \R$ be a function which is
differentiable on a possibly unbounded open interval $I\subset \R$ and 0
elsewhere.  Suppose $\odd(\fOE_{n-1}(f)\cup \fOE_n(f))=\fUE_n(g)$ for
some function $g$.  Then $f$ and $g$ have the form
\begin{align}
f(x) &\propto (1-rx)^{-n},\\
g(x) &\propto (1-rx)^{-2n+1}
\end{align}
for some $r$ (possibily $\infty$) with $1/r\not\in I$.
\end{thm}

\begin{proof}
The only tricky aspect of this case is that the determinant we must
analyze is no longer of the form to which Theorem \ref{thm:pOE=OE}
applies; to be precise, we need
\[
\det(F_j(x_{2i+2})-F_j(x_{2i}))_{0\le i<n}
\]
to have orthogonal ensemble form.  But this determinant is clearly equal to
the determinant of the block matrix
\[
\pmatrix
(1)                   &(0)_{0\le i<n}\\
(F_j(x_0))_{0\le j<n} &(F_j(x_{2i+2})-F_j(x_{2i}))_{0\le i,j<n}.
\endpmatrix
\]
Adding the first column to the other columns, we can then apply Theorem
\ref{thm:pOE=OE}, and argue as above.
\end{proof}

We finally consider a fifth possibility for decimation.  Recall that for
the circular ensemble results cited above, while there was a {\it local}
notion of order, there was no notion of largest.  This suggests that
we consider the ensemble derived by choosing randomly between $\odd(M)$
and $\even(M)$.  More precisely, for an ensemble with an even
number of variables, we define $\alt(M)$ to be $\even(M)$ with probability
$1/2$ and $\odd(M)$ with probability $1/2$.

\begin{thm}\label{thm3.6}
Fix an integer $n\ge 2$.  Let $f:\R\to \R$ be a function which is
differentiable on a possibly unbounded open interval $I\subset \R$ and 0
elsewhere.  Suppose $\alt(\fOE_n(f)\cup \fOE_n(f))=\fUE_n(g)$ for some
function $g$.  Then up to a linear transformation of
variables, $f$ and $g$ have the form
\begin{align}
f &= (1+x^2)^{-(n+1)/2}\\
g &= (1+x^2)^{-n}.
\end{align}
\end{thm}

\begin{proof}
Consider the determinants associated to $\even(\fOE_n(f)\cup \fOE_n(f))$
and $\odd(\fOE_n(f)\cup \fOE_n(f))$.  Up to cyclic shift, only one column
differs between the two determinants, thus allowing us to express
their sum as a determinant.  When $n$ is even, the `special' column takes
the form
\[
(F_j(x_0)+F_j(x_{2n-1})-F_j(I))_{0\le j<n};
\]
here $F_j(I)=\int_{x\in I} x^j f(x)$.  Taking appropriate linear
combinations, we obtain the determinant
\[
\det(F_j(x_i)-F_j(I)/2)_{0\le i,j<n}.
\]
When $n$ is odd, the special column takes the form
\[
(F_j(x_{2n-1})-F_j(x_0)-F_j(I))_{0\le j<n};
\]
this leads (up to sign) to the $n+1\times n+1$ block determinant
\[
\det\pmatrix
0&(F_j(I))_{0\le j<n}\\
1&(F_j(x_i))_{0\le i,j<n}
\endpmatrix.
\]

We first analyze the case $n$ odd.  In this case, the usual theory tells
us that there exist polynomials $p_j(x)$ and $q(x)$ of degree at most 
$n-1$ with
\[
F_j(x)-C_j F_0(x) = p_j(x)/q(x)
\]
for all $j$, with $C_j$ as above.  Now, evaluating this at an endpoint
of $I$, we find that the polynomials $p_j$ must have a common root
(possibly $\infty$).  In particular, it follows that $f$ must satisfy
the conditions of Theorem \ref{thm:OEOE.2}.  On the other hand, we find
that
\[
x^j f(x) - C_j f(x) = {d\over dx} {p_j(x)\over q(x)}
\]
for each $j$; in particular, $f(x)$ must be a rational function.
We therefore have the following possibilities to consider:
\begin{align}
f(x) &= x^\alpha (1-x)^\beta,\quad \alpha,\beta\in \N,\\
f(x) &= (1+x^2)^{-\alpha},\quad \alpha\in \N, \alpha>n/2,
\end{align}
on $(0,1)$ and $\R$ respectively.  In the first case, we find that each
\[
F_j(x)-C_j F_0(x)
\]
is a polynomial of degree $j+2$.  In particular, $F_{n-1}(x)-C_{n-1}F_0(x)$
is a polynomial of degree $n+1$, contradicting the bound on $\deg(p_j(x))$.

In the second case, we observe that
\[
F_j(x)-C_j F_0(x) = r_j(x) (1+x^2)^{1-\alpha}
\]
for polynomials $r_j(x)$ of degree $j-1$.  In particular, we find that
$p_1(x)/q(x)\propto (1+x^2)^{1-\alpha}$, implying, since $\alpha>n/2>1$,
that
\[
q(x) \propto (1+x^2)^{\alpha-1}
\]
Since $\deg(q)\le n-1$, we have $n/2<\alpha\le (n+1)/2$, the only integral
solution of which is $\alpha=(n+1)/2$.  In this case, the relevant degree
bounds all hold, and thus the determinant is indeed of the correct form,
giving $g(x)\propto (1+x^2)^{1-2\alpha}=(1+x^2)^{-n}$ as required.

For $n$ even, we must have polynomials $p_i$ of degree at most $n-1$ with
\[
p_i(x) (F_j(x)-C'_j) = p_j(x) (F_i(x)-C'_i),
\]
where we write $C'_j$ for $F_j(I)/2$.  We can rewrite this as:
\[
p_0(x)(F_j(x)-C_j F_0(x))
=
(p_j(x)-C_j p_0(x))(F_0(x)-C'_0),
\]
using the fact that $C_j=C'_j/C'_0$.  For $n>2$, we conclude that $f(x)$
must satisfy the conditions of Theorem \ref{thm:OEOE.2}.  Now, if the
endpoints of $I$ are different, then we find, since $F_0(x)-C'_0=\pm C'_0$
at both endpoints, that each $p_j(x)-C_j p_0(x)=0$ at both endpoints.  But
this causes the polynomials to be linearly dependent, a contradiction.
On the other hand, in the other cases, we know that $C_1=0$ and $p_1\propto
1$.  In both cases, we obtain from the identity for $F_1(x)-C_1 F_0(x)$
a differential equation for $p_0(x)$.  For $e^{-x^2/2}$, no polynomial
solution to the equation exists.  For $(1+x^2)^{-\alpha}$, we can find
an explicit power series solution to the equation, and find that
a polynomial solution exists only when $\alpha$ is half-integral, when
the solution has degree $2\alpha-2$.  As above, this leaves only one
possibility for $\alpha$, namely $\alpha=(n+1)/2$, as required.

It remains to consider $n=2$.  Here we can twice differentiate the equation
\[
p_0(x) (F_1(x)-C'_1) = p_1(x) (F_0(x)-C'_0)
\]
(with $p_0$ and $p_1$ linear) to deduce that
\[
f'(x)/f(x) = p(x)/q(x)
\]
with $\deg(p)\le 1$, $\deg(q)\le 2$, and $\deg(xp+3q)\le 1$.
So up to LFT, $f(x)$ must have one of the forms
\begin{align}
f(x) &= x^\alpha,\quad \alpha \ge -3/2 \\
f(x) &= e^{-x}\\
f(x) &= (1+x^2)^{-3/2}.\\
\end{align}
(note that if we exchange $0$ and $\infty$ in the first case, we replace
$\alpha$ with $-3-\alpha$, justifying our restriction on $\alpha$.)  In the
first case, if $\alpha\ne -1$, the following:
\[
{x^{\alpha+2}-D_1\over x^{\alpha+1}-D_0}
\]
must be a linear polynomial for suitable constants $D_1$
and $D_0$ respectively proportional to $C'_1$ and $C'_0$ (and thus $D_0\ne
0$).  We deduce therefore that $\alpha=0$.  But then we readily determine
that only the empty interval satisfies the requirements.  For
$x^{-1}$ and $e^{-x}$, there are not even appropriate choices for $D_0$ and
$D_1$ (since both $\log(x)$ and $e^{-x}$ are transcendental functions).
Finally, for the third choice, we readily verify that decimation indeed
works as required.
\end{proof}

We now turn our attention to decimations of single orthogonal ensembles.
We have, quite simply:

\begin{thm}\label{thm3.7}
Fix an integer $n\ge 2$.  For any functions $f$ and $g$ with $f$
differentiable on a possibly unbounded open interval
$I\subset \R$ and 0 elsewhere, each of the following pairs
of statements is equivalent:
\begin{align}
\even(\fOE_{2n}(f)\cup\fOE_{2n+1}(f))=\fUE_{2n}(g)
&\text{\ \ and\ \ }
\even(\fOE_{2n+1}(f))=\fSE_n((g/f)^2)\label{3.85}\\
\even(\fOE_{2n}(f)\cup\fOE_{2n}(f))=\fUE_{2n}(g)
&\text{\ \ and\ \ }
\even(\fOE_{2n}(f))=\fSE_n((g/f)^2)\label{3.86}\\
\odd(\fOE_{2n}(f)\cup\fOE_{2n}(f))=\fUE_{2n}(g)
&\text{\ \ and\ \ }
\odd(\fOE_{2n}(f))=\fSE_n((g/f)^2)\label{3.87}\\
\odd(\fOE_{2n-1}(f)\cup\fOE_{2n}(f))=\fUE_{2n}(g)
&\text{\ \ and\ \ }
\odd(\fOE_{2n-1}(f))=\fSE_n((g/f)^2) \label{3.88}\\
\alt(\fOE_{2n}(f)\cup\fOE_{2n}(f))=\fUE_{2n}(g)
&\text{\ \ and\ \ }
\alt(\fOE_{2n}(f))=\fSE_n((g/f)^2). \label{3.89}
\end{align}
\end{thm}

\begin{proof}
Consider, for instance, $\even(\fOE_{2n+1}(f))$.  Once
we integrate along the largest, 3rd largest, etc. variables and
do some simplification, the resulting matrix has columns
$(F_j(x_{2i+1}))_{0\le j\le 2n}$ and $(x_{2i+1}^j f(x_{2i+1}))_{0\le j\le 2n}$,
with the last column given by $(F_j(b))_{0\le j\le 2n}$.
In particular, we note that aside from the last, constant, column,
the columns come in pairs, one the derivative of the other.  Thus
the determinant is essentially of the form considered in Theorem
\ref{thm:pSE=SE}.  In particular, it has a symplectic ensemble form
if and only if the determinant
\[
\det(F_j(x_{i+1}))_{0\le i,j\le 2n}
\]
(of which our determinant is a derivative) has an orthogonal ensemble form.
But by the proof of Theorems \ref{thm:OEOE.1} and \ref{thm:OEOE.2}, this
precisely what we needed to show.  (The statement about the resulting weight
functions is straightforward.)  Similar arguments apply for the
remaining equivalences.
\end{proof}

\section{Random matrix applications}

In random matrix applications $f$, $g$ and $(g/f)^2$ must be (up to the
scale of $x$) one of the four classical forms (\ref{clw}) and
(\ref{clw1}). So specializing Theorems \ref{thm:OEOE.2}--\ref{thm3.5}
we can read off for which of the classical forms the statement of
Theorem \ref{thm3.7} is valid.

\begin{thm}\label{thm4.1}
Restricting attention to the classical weights (\ref{clw}) and
(\ref{clw1}), the statement (\ref{3.85}) holds for
\begin{equation}\label{4.1a}
(f,g) = \left \{ \begin{array}{l}
(e^{-x^2/2}, e^{-x^2}) \\
(x^{(a-1)/2} e^{-x/2}, x^ae^{-x}), \: \: x>0\\
(x^{(a-1)/2}(1-x)^{(b-1)/2}, x^a(1-x)^b),\:\: 0<x<1 \\
((1+x^2)^{-(\alpha+1)/2}, (1+x^2)^{-\alpha}), 
\end{array} \right.
\end{equation}
the statement (\ref{3.86}) holds for
\begin{equation}\label{4.1b}
(f,g) = \left \{ \begin{array}{l} (e^{-x/2}, e^{-x}) \\
((1-x)^{(a-1)/2}, (1-x)^a), \:\: 0<x<1\end{array} \right.
\end{equation}
while (\ref{3.87}) is valid for the particular pair of Jacobi weights
\begin{equation}\label{4.1c}
(f,g) = (x^{(a-1)/2}, x^a)
\end{equation}
and (\ref{3.88}) is valid for the particular pair of Jacobi weights
\begin{equation}\label{4.1d}
(f,g)=(1,1).
\end{equation}
\end{thm}

Because the weights in Theorem \ref{thm4.1} occur in the matrix
ensembles listed in the Introduction, the theorems of
Section \ref{s3} imply inter-relationships between the different
ensembles. 

\begin{thm}
The following relations hold between the above matrix ensembles under
decimation, for all $n>0$:
\begin{align}
\even(\GOE_{2n+1}) &= \GSE_n\\
\even(\Symm(n;\C)) &= \Anti(n;\C)\\
\even(\Mat(2p+1,2q+1;\R)) &= \Mat(p,q;\H)\\
\even({\rm Beta}(2p_1+1,2p_2+1,2q+1;\R)) &=
{\rm Beta}(p_1,p_2,q;\H)\\
\noalign{\medskip}
\even(\GOE_n\cup \GOE_{n+1}) &= \GUE_n\\
\even(\Symm(n;\C)\cup \Symm(n;\C)) &= \Mat(n,n;\C)\\
\even(\Symm(n;\C)\cup \Symm(n+1;\C)) &= \Mat(n+1,n;\C)\\
\even(\Mat(p,q;\R)\cup \Mat(p+1,q+1;\R)) &= \Mat(p,q;\C)\\
\even({\rm Beta}(p_1,p_2,q;\R) \cup
{\rm Beta}(p_1+1,p_2+1,q+1;\R)) &= {\rm Beta}(p_1,p_2,q;\C)
\end{align}
\end{thm}

\begin{rems}
It would be very nice to have a direct, matrix-theoretic, proof of {\em
any} of the above relations.
\end{rems}

\begin{rems}
There are actually a few more relations, all of which follow from
the above together with the relation
\[
\Mat(n+1,n;\R) = \Symm(n;\C).
\]
Again, a matrix-theoretic proof of this would be nice.
\end{rems}

We now turn our attention to the implications of Theorem \ref{thm4.1}
with respect to gap probabilities.
In circular ensemble theory the results (\ref{0.9}) and (\ref{0.10})
were shown \cite{Dy62c} to imply inter-relationships between the
probability of an eigenvalue free region amongst the various symmetry
classes. With $E^{(\beta)}(p;J;n)$ denoting the probability that, for the
ensembles COE$_n$ ($\beta = 1$), CUE$_n$ ($\beta = 2$) and CSE$_n$
($\beta = 4$), there are exactly $p$ eigenvalues in the interval $J$,
the inter-relationships are
\begin{align}\label{cdy}
E^{(2)}(0;(0,s);n) & = E^{(1)}(0;(0,s);n) \Big (
E^{(1)}(0;(0,s);n) + E^{(1)}(1;(0,s);n) \Big ) \nonumber \\
E^{(4)}(p;(0,s);n) & = E^{(1)}(2p;(0,s);2n) + {1 \over 2}
E^{(1)}(2p-1;(0,s);2n) + {1 \over 2} E^{(1)}(2p+1;(0,s);2n)
\end{align}
where 
\begin{equation}\label{neg}
E^{(\beta)}(p;J;n) := 0, \quad \text{for} \quad p<0.
\end{equation}
Similar inter-relationships
between gap probabilities, but now with the eigenvalue free interval
including an endpoint of the support of the interval, can be deduced from
the pairs of statements of Theorem \ref{thm3.7}.

\begin{thm}
Let $E^{(\beta)}(p;J;g;n)$ denote the probability that, for the ensembles
$\fOE_n(g)$ ($\beta = 1$), $\fUE_n(g)$ ($\beta = 2$) and $\fSE_n(g)$
($\beta = 4$) the interval $J$ contains exactly $n$ eigenvalues. The
statements (\ref{3.85}) imply
\begin{eqnarray}
E^{(2)}(0;J;g;2n) & = & E^{(1)}(0;J;f;2n)E^{(1)}(0;J;f;2n+1) +
E^{(1)}(0;J;f;2n)E^{(1)}(1;J;f;2n+1) \nonumber \\
&& + E^{(1)}(0;J;f;2n+1)E^{(1)}(1;J;f;2n) \nonumber \\
E^{(4)}(p;J;(g/f)^2;n) & = & E^{(1)}(2p;J;f;2n+1) +
E^{(1)}(2p+1;J;f;2n+1) \label{3.85'}
\end{eqnarray}
for $J = (-\infty, -s)$ or $(s,\infty)$; the statements (\ref{3.86}) imply
\begin{eqnarray}\label{3.86'}
E^{(2)}(0;(-\infty,-s);g;2n) & = &
\Big (E^{(1)}(0;(s,\infty);f;2n)\Big )^2 + 2
E^{(1)}(0;(s,\infty);f;2n)E^{(1)}(1;(s,\infty);f;2n) \nonumber \\
E^{(4)}(p;(s,\infty);(g/f)^2;n) & = &
E^{(1)}(2p;(s,\infty);f;2n) + E^{(1)}(2p+1;(s,\infty);f;2n)
\end{eqnarray}
and
\begin{eqnarray}\label{3.86''}
E^{(2)}(0;(-\infty,-s);g;2n) & = &
\Big (E^{(1)}(0;(-\infty,-s);f;2n)\Big )^2 \nonumber \\
E^{(4)}(p;(-\infty,-s);(g/f)^2;n) & = &
E^{(1)}(2p;(-\infty,-s);f;2n) + E^{(1)}(2p-1;(-\infty,-s);f;2n);
\end{eqnarray}
the statements (\ref{3.87}) imply
\begin{eqnarray}\label{3.87'}
E^{(2)}(0;(s,\infty);g;2n) & = &
\Big (E^{(1)}(0;(s,\infty);f;2n)\Big )^2 \nonumber \\
E^{(4)}(p;(s,\infty);(g/f)^2;n) & = &
E^{(1)}(2p;(s,\infty);f;2n) + E^{(1)}(2p-1;(s,\infty);f;2n)
\end{eqnarray}
and
\begin{eqnarray}\label{3.87''}
\lefteqn{
E^{(2)}(0;(-\infty,-s);g;2n) } \nonumber \\&& 
= \Big (E^{(1)}(0;(-\infty,-s);f;2n)\Big )^2 
+ 2 E^{(1)}(0;(-\infty,-s);f;2n)E^{(1)}(1;(-\infty,-s);f;2n)\nonumber \\
&& E^{(4)}(p;(-\infty,-s);(g/f)^2;n)  = 
E^{(1)}(2p;(-\infty,-s);f;2n) + E^{(1)}(2p+1;(-\infty,-s);f;2n);
\end{eqnarray}
the statements (\ref{3.88}) imply
\begin{eqnarray}\label{3.88'}
E^{(2)}(0;J;g;2n) & = & E^{(1)}(0;J;f;2n) E^{(1)}(0;J;f;2n-1)
\nonumber \\
E^{(4)}(p;J;(g/f)^2;n) & = & E^{(1)}(2p;J;f;2n-1) + E^{(1)}(2p-1;J;f;2n-1)
\end{eqnarray}
for $J=(-\infty,-s)$ or $(s,\infty)$, while the statements (\ref{3.89})
imply the relations (\ref{cdy}) with $n$ replaced by $2n$ in the first
equation and $(0,s)$ replaced throughout by $J$, 
 $J=(-\infty,-s)$ or $(s,\infty)$.
\end{thm}

\begin{proof}
We will consider only the deductions from (\ref{3.85}), as the other cases
are similar. Let $J$ be a single interval which includes an endpoint of
the support of $f$ and $g$. From the first statement in (\ref{3.85}) we
see that the event of a sequence of eigenvalues from $\fUE_{2n}(g)$ 
not being contained in $J$ occurs in three ways relative to the
ensemble $\fOE_{2n}(f) \cup \fOE_{2n+1}(f)$: (i) the eigenvalues from
$\fOE_{2n}(f)$ and those from $\fOE_{2n+1}(f)$ are not contained in $J$;
or (ii) one eigenvalue from $\fOE_{2n+1}(f)$ is contained in $J$ and no
eigenvalue from $\fOE_{2n}(f)$ is contained in $J$ (note that the
one eigenvalue must be either the largest (smallest) eigenvalue
 when $J$ contains the right (left) hand end point); or (iii)
one eigenvalue from $\fOE_{2n}(f)$ is contained in $J$ and no
eigenvalue from $\fOE_{2n+1}(f)$  is contained in $J$. This gives the
first equation in (\ref{3.85'}). From the second statement in
(\ref{3.85}) we see that the event of a sequence of eigenvalues
from $\fSE_n((g/f)^2)$ containing $p$ eigenvalues
in $J$ can occur in two ways relative to $\fOE_{2n+1}(f)$:
(i) there are $2p$ eigenvalues from  $\fOE_{2n+1}(f)$ in $J$; or (ii) 
there are $2p+1$ eigenvalues from  $\fOE_{2n+1}(f)$ in $J$ (of which
$p+1$ are integrated out in forming even($\fOE_{2n+1}(f))$). This
implies the second equation in (\ref{3.85'}).
\end{proof}

Recalling (\ref{neg}) we see that in the case $p=0$ the equations
(\ref{3.86''}), (\ref{3.87'}) and (\ref{3.88'}) give particularly simple
inter-relationships between the $E^{(\beta)}(0;\dots)$. In fact referring
back to Theorem \ref{thm4.1} for the permissable pairs
$(f,g)$ in these cases
it is a simple
exercise in changing variables to compute the $E^{(\beta)}(0;\dots)$
in terms of elementary functions. Recalling
$$
E^{(\beta)}(0;J;w;n) := {1 \over C}
\int_{\bar{J}}dx_0 \cdots \int_{\bar{J}} dx_{n-1} \,
\prod_{l=0}^{n-1} w(x_l) \prod_{0 \le j < k \le n-1} |x_k - x_j|^\beta,
$$
where $\bar{J}=(-\infty,\infty) - J$ and $C$ is such that
$E^{(\beta)}(0;\emptyset;w;n) = 1$ we find
\begin{align*}
E^{(1)}(0;(0,s);e^{-x/2};n) & = e^{-sn/2} \\
E^{(2)}(0;(0,s);e^{-x};n) & = E^{(4)}(0;(0,s);e^{-x};n) = e^{-sn} \\
E^{(1)}(0;(0,s);(1-x)^{(a-1)/2};n) & =
E^{(1)}(0;(1-s,1);x^{(a-1)/2};n) = (1-s)^{n(n+a)/2} \\
E^{(2)}(0;(0,s);(1-x)^a;n) & = E^{(2)}(0;(1-s,1);x^a;n) =
(1-s)^{n(a+n)} \\
E^{(4)}(0;(0,s);(1-x)^{a+1};n) & = E^{(4)}(0;(1-s,1);x^{a+1};n) =
(1-s)^{2n^2+na}
\end{align*}
(in the first two cases the weight functions are restricted to $x>0$, while
in the remaining cases $0<x<1$). The equations (\ref{3.86''}), (\ref{3.87'})
and (\ref{3.88'}) for $p=0$ can be checked immediately.

The pairs of equations (\ref{3.85'}), (\ref{3.86'}) and (\ref{3.87''})
contain $E^{(1)}(1;\dots)$ as well as $E^{(\beta)}(0;\dots)$. In
the equations (\ref{3.86'}) and (\ref{3.87''}) the dependence on
$E^{(1)}(1;\dots)$  can be eliminated. 
Noting from Theorem \ref{thm4.1} the allowed pairs
$(f,g)$
for the validity of  equations (\ref{3.86'}) and (\ref{3.87''})
the following result is obtained.

\begin{prop}
For $(f,g) = (e^{-x/2}, e^{-x})$, $(x>0)$, and $J=(s,\infty)$,
or for $(f,g)=((1-x)^{(a-1)/2},(1-x)^a)$, $(0<x<1)$, and 
$J=(1-s,1)$ we have
\begin{equation}\label{4.1}
E^{(4)}(0;J;(g/f)^2;n) = {1 \over 2}
\Big ( E^{(1)}(0;J;f;2n) +
{E^{(2)}(0;J;g;2n) \over E^{(1)}(0;J;f;2n)} \Big ).
\end{equation}
\end{prop}

In the scaled $n \to \infty$ limit, as appropriate for the particular
choice of weight function in (\ref{4.1a}), the pair of equations
(\ref{3.85}) also imply an equation of the form (\ref{4.1}).
First consider the Gaussian ensembles with the scaling \cite{Fo93a}
$$
x \mapsto (2n)^{1/2} + {x \over 2^{1/2} n^{1/6}},
$$
which corresponds to studying the distribution of the eigenvalues at the
(soft) edge of the leading order support of the spectrum. Defining
\begin{eqnarray*}
E_{\rm soft}^{(1)}(p;(s,\infty)) & := &
\lim_{n \to \infty} E^{(1)}\Big (p, ((2n)^{1/2} + {s \over 2^{1/2}
n^{1/6}},\infty); e^{-x^2/2};n \Big ) \\
E_{\rm soft}^{(2)}(p;(s,\infty)) & := &
\lim_{n \to \infty} E^{(2)}\Big (p, ((2n)^{1/2} + {s \over 2^{1/2}
n^{1/6}},\infty); e^{-x^2};n \Big ) \\
E_{\rm soft}^{(4)}(p;(s,\infty)) & := &
\lim_{n \to \infty} E^{(4)}\Big (p, ((2n)^{1/2} + {s \over 2^{1/2}
n^{1/6}},\infty); e^{-x^2};n/2 \Big ),
\end{eqnarray*}
(the existence of these limits is known from explicit calculation
\cite{TW96}; see below).  The equations (\ref{3.85'}) imply:

\begin{prop}\label{thm4.5}
For the scaled infinite Gaussian ensembles at the soft edge
\begin{eqnarray}
E_{\rm soft}^{(4)}(0;(s,\infty)) & = & {1 \over 2}
\Big ( E_{\rm soft}^{(1)}(0;(s,\infty)) +
{E_{\rm soft}^{(2)}(0;(s,\infty)) \over  E_{\rm soft}^{(1)}(0;(s,\infty))}
\Big )  \label{16a}\\
E_{\rm soft}^{(1)}(1;(s,\infty)) & = &
E^{(4)}_{\rm soft}(0;(0,\infty)) - E_{\rm soft}^{(1)}(0;(s,\infty)) \label{16b}
\end{eqnarray}
\end{prop}

As alluded to above,
 the $E_{\rm soft}^{(\beta)}(0;(s,\infty))$ are known exactly from
the work of Tracy and Widom \cite{TW96}. To present these results, let
$q(s)$ denote the solution of the particular Painlev\'e II equation
$$
q'' = sq + 2 q^3
$$
which satisfies the boundary condition
$
q(s) \sim{\rm Ai}(s)$  as $s \to \infty$.
Then we have
\begin{eqnarray}\label{b4}
E_{\rm soft}^{(2)}(0;(s,\infty)) 
& = & \exp \Big ( - \int_s^\infty (t-s) q^2(t) \, dt \Big )
\nonumber \\
\Big ( E_{\rm soft}^{(1)}(0;(s,\infty)) \Big )^2 & = & 
E_{\rm soft}^{(2)}(0;(s,\infty))
\exp \Big ( - \int_s^\infty q(t) \, dt \Big ) \nonumber \\
\Big ( E_{\rm soft}^{(4)}(0;(s,\infty)) \Big )^2 & = & 
E_{\rm soft}^{(2)}(0;(s,\infty))
\cosh^2 \Big ( {1 \over 2} \int_s^\infty q(t) \, dt \Big ),
\end{eqnarray}
(in \cite{TW96} $E_{\rm soft}^{(4)}$ is defined with $s \mapsto
s/2^{1/2}$ relative to our definition). The equation (\ref{16a})
is immediately seen to be satisfied, while the second equation gives
\begin{equation}\label{16a'}
\Big ( E_{\rm soft}^{(1)}(1;(s,\infty)) \Big )^2  
=  E_{\rm soft}^{(2)}(0;(s,\infty))
\sinh^2 \Big ( {1 \over 2} \int_s^\infty q(t) \, dt \Big ).
\end{equation}

Next consider the scaled limit at an edge for which the weight function
is strictly zero on one side. For the classical ensembles this occurs in
the Laguerre and Jacobi case; for definiteness consider the Laguerre
case. The appropriate scaling is \cite{Fo93a}
$$
x \mapsto {x \over 4n},
$$
and we define
\begin{eqnarray*}
E^{(1)}_{\rm hard}(p;(0,s);(a-1)/2) & := &
\lim_{n \to \infty} E^{(1)}(p;(0,s/4n);x^{(a-1)/2}e^{-x};n) \\
E^{(2)}_{\rm hard}(p;(0,s);a) & := &
\lim_{n \to \infty}E^{(2)}(p;(0,s/4n);x^ae^{-x};n) \\
E^{(4)}_{\rm hard}(p;(0,s);a+1) & := &
\lim_{n \to \infty}E^{(4)}(p;(0,s/4n);x^{a+1}e^{-x};n/2)
\end{eqnarray*}
(the existence of these limits for general $a>-1$ can be deduced from
the existence of the $k$-point distributions in the same scaled limits
\cite{NF95}). Use of (\ref{3.85}) then gives the analogue of
Proposition \ref{thm4.5} for the hard edge.

\begin{prop}
For the scaled infinite Laguerre ensembles at the hard edge
\begin{eqnarray}
E_{\rm hard}^{(4)}(0;(0,s);a+1) & = & {1 \over 2}
\Big ( E_{\rm hard}^{(1)}(0;(0,s);(a-1)/2) +
{E_{\rm hard}^{(2)}(0;(0,s);a) \over  
E_{\rm hard}^{(1)}(0;(0,s);(a-1)/2)}
\Big )  \label{17a}\\
E_{\rm hard}^{(1)}(1;(0,s);(a-1)/2) & = &
E^{(4)}_{\rm hard}(0;(0,s);a+1) - E_{\rm hard}^{(1)}(0;(0,s);
(a-1)/2). \label{17b}
\end{eqnarray}
\end{prop}

Exact Pfaffian formulas are known for $E_{\rm hard}^{(1)}(0;(0,s);(a-1)/2)$
and $E^{(4)}_{\rm hard}(0;(0,s);a+1)$ in the case $a$ an odd positive
integer \cite{NF97i}, while $E_{\rm hard}^{(2)}(0;(0,s);a)$ can then be 
expressed as a determinant \cite{FH94} (the dimension of the
Pfaffians and the determinants are proportional to $a$), although
(\ref{17a}) is not a natural consequence of these formulas. There are
also multiple integral expressions for the same expression
\cite{Fo93c}, but again they do not naturally satisfy (\ref{17a}).

\section{Distribution functions for superimposed spectra}
In general, for a symmetric PDF $p(x_0,\dots,x_{n-1})$ the $k$-point
distribution function $\rho_k$ is defined by
\begin{equation}\label{5.1}
\rho_k(x_0,\dots,x_{k-1}) :=
n(n-1)\cdots(n-k+1) \int_{(-\infty,\infty)^{n-k}}
p(x_0,\dots,x_{n-1})dx_{k} \cdots dx_{n-1}.
\end{equation}
In this section we take up the task of computing $\rho_k$ for
even$(M)$, odd$(M)$, alt$(M)$ with $M = \fOE_n(f) \cup
\fOE_n(f)$ and even$(M)$, odd$(M)$ with $M = \fOE_n(f) \cup
\fOE_{n+1}(f)$.

For $M = \fOE_n(f) \cup \fOE_n(f)$ write
\begin{eqnarray}\label{c1}
D^{\rm even(M)}(x_0,\dots,x_{n-1}) & := &
\det( F_{j}(x_i) - F_{j}(I))_{0\le i,j <n} \nonumber \\
D^{\rm odd(M)}(x_0,\dots,x_{n-1}) & := &
\det( F_{j}(x_i))_{0\le i,j <n} \nonumber \\
D^{\rm alt(M)}(x_0,\dots,x_{n-1}) & := &
\det( F_{j}(x_i) - {1 \over 2}
F_{j}(I) )_{0\le i,j <n} \quad (n \: {\rm even}) \nonumber \\
D^{\rm alt(M)}(x_0,\dots,x_{n-1}) & := &
\det \left ( \begin{array}{cc}
0 & (F_j(I))_{0\le j <n} \nonumber \\
(1)_{1\le i < n-1} & (F_j(x_i))_{0\le i,j < n} \end{array}
\right ) \quad (n \: {\rm odd})
\end{eqnarray}
and for $M = \fOE_n(f) \cup \fOE_{n+1}(f)$ let
\begin{eqnarray}\label{c2}
D^{\rm even(M)}(x_0,\dots,x_{n-1}) & := &
\det \left ( \begin{array}{c}
(F_{j}(x_i))_{\substack{0 \le i < n\\0 \le j < n+1}} \\
(F_{j}(I))_{0 \le j < n+1}  \end{array} \right ) \nonumber \\
D^{\rm odd(M)}(x_0,\dots,x_{n-1}) & := &
\det \left ( \begin{array}{c}
(1)_{0 \le j < n+1} \\ 
(F_j(x_i))_{\substack{0 \le i < n \\ 0 \le j < n+1}} \end{array} \right )
\end{eqnarray}

In each case, workings contained in the proofs of Theorems
\ref{thm:OEOE.1} and \ref{thm3.6} show (after relabelling the
coordinates) that the PDF is proportional to
\begin{equation}\label{S}
\prod_{i=0}^{n-1} f(x_i) \Delta(x_0,\dots,x_{n-1})
D(x_0,\dots,x_{n-1})
\end{equation}
for $D$ as specified.
Now introduce a set of functions $\{ \eta_j(x) \}_{0\le j < n}$
such that
$$
D(x_0,\dots,x_{n-1}) \propto \det(\eta_j(x_i))_{0\le i,j < n}
$$
and a set of monic polynomials $\{ q_j(x)\}_{0\le j < n}$, $q_j$ of
degree $j$, such that the biorthogonality property
$$
\int_{-\infty}^\infty f(x) q_i(x) \eta_j(x) \, dx =
\delta_{i,j}
$$
holds (assuming such biorthogonal families exist). The $k$-point distribution
can be expressed in terms of these functions.

\begin{lem}
For the PDF (\ref{S}) and  $\{ \eta_j(x) \}_{0\le j < n}$,
$\{ q_j(x)\}_{0\le j < n}$ specified as above, we have
\begin{equation}\label{S8}
\rho_k(x_0,\dots,x_{k-1}) = \prod_{j=0}^{k-1} f(x_i) 
\det \Big ( \sum_{l=0}^{n-1} q_l(x_i) \eta_l(x_j)
\Big )_{0\le i,j < k}.
\end{equation}
\end{lem}

\begin{proof}
 From the definitions of $\{ \eta_j(x) \}_{0\le j < n}$ and
$\{ q_j(x)\}_{0\le j < n}$ we see that (\ref{S}) is
proportional to
$$
\prod_{i=0}^{n-1} f(x_i) \det (q_j(x_i))_{0 \le i,j < n}
\det (\eta_j(x_i))_{0\le i,j < n}.
$$
The biorthogonal property allows the integrations required by
the definition (\ref{5.1}) to be computed to give (\ref{S8}).
\end{proof}

\begin{rem} Suppose for some $\{\xi_j(x)\}_{j=0,\dots,n-1}$ we can write
$$
D(x_0,\dots,x_{n-1}) \propto \det (\xi_j(x_i) )_{0 \le i,j < n-1}.
$$
It is easy to show \cite{MMN98,Bor99}
that sufficient conditions for the existence of the
biorthogonal sets is that
$$
\det \Big ( \int_{-\infty}^\infty f(x) x^i \xi_j(x) \, dx \Big )_{
0 \le i,j < p} \ne 0
$$
for $p=0,\dots,n-1$.
\end{rem}

For $f$ a classical weight and so of the form (\ref{clw}) or 
(\ref{clw1}), the biorthogonal functions can be computed explicitly. This
is possible because of the following special property of the classical
weights and their corresponding orthogonality \cite{AFNV99}.

\begin{lem}\label{n}
Consider the pairs $(f,g)$ of classical weight functions (\ref{4.1a}).
Let $\{p_j(x)\}_{j=0,1,\dots}$ be the set of monic orthogonal polynomials,
$p_j(x)$ of degree $j$ corresponding to the weight function $g$,
let $(p_k,p_k)_2$ denote their normalization with respect to
integration over the measure $g(x) dx$,
and define
$\gamma_k$ so that
$$
\gamma_k (p_k,p_k)_2 = \left \{
\begin{array}{l} 1 \\ {1 \over 2} \\ {1 \over 2}(2k+2+a+b) \\ 
\alpha - k - 1 \end{array} \right.
$$
in the four cases respectively. With
$$
\mbox{\boldmath$n$} \,:= {1 \over f(x)} {d \over dx} 
{g(x) \over f(x)}, \qquad c_k := \gamma_k (p_k,p_k)_2
(p_{k+1},p_{k+1})_2
$$
we have 
$$
\mbox{\boldmath$n$} \, p_k(x) = - {c_k \over (p_{k+1},p_{k+1})_2} p_{k+1}(x)
+ {c_{k-1} \over (p_{k-1},p_{k-1})_2} p_{k-1}(x).
$$
\end{lem}

\begin{proof}
This is a simple consequence of the property \cite{AV95}
$$
(\phi,\mbox{\boldmath$n$} \, \psi)_2 = - 
(\mbox{\boldmath$n$} \, \phi, \psi)_2.
$$
\end{proof}

As a consequence, the determinant formulas 
(\ref{c1}) and (\ref{c2}) for $D(x_0,\dots,x_{n-1})$  can be simplified.
Let 
\begin{equation}\label{u}
[u(x)]_j := \sum_{l=j}^\infty {(u(x),p_l(x))_2 \over (p_l,p_l)_2}p_l(x)
=u(x) - \sum_{l=0}^{j-1} {(u(x),p_l(x))_2 \over (p_l,p_l)_2}p_l(x)
\end{equation}
and define
$$
\begin{array}{ll}
r_{j}^{(1)}(x) = \Big [{f(x) \over g(x)} \int_x^\infty f(t) \, dt
\Big ]_{j} &
r_{j}^{(2)}(x) = \Big [{f(x) \over g(x)}
\int_{-\infty}^x f(t) \, dt \Big ]_{j} \\
r_{j}^{(3)}(x) = 
\Big [{f(x) \over g(x)} \Big (
\int_{-\infty}^x f(t) \, dt 
-{1 \over 2} \int_{-\infty}^\infty f(t) \, dt \Big ) \Big ]_{j}
& r_{j}^{(4)}(x) =
\Big [{f(x) \over g(x)} \Big ]_{j} \\
r_j^{(5)}(x) = r_{j-1}^{(4)}(x) - {(r_{j-1}^{(4)},r_{j-1}^{(4)})_2 \over
(r_{j-1}^{(4)},r_{j-1}^{(2)})_2} r_{j-1}^{(2)}(x) &
\end{array}
$$
Then by adding appropriate linear combinations of the columns in
the determinant formulas (\ref{c1}) and (\ref{c2}), and making use
of Lemma \ref{n}  we readily find for $M = \fOE_n(f) \cup \fOE_n(f)$
\begin{align}
D^{\text{even}(M)}(x_0,\dots,x_{n-1}) & \propto
\prod_{i=1}^{n-1} {g(x_i) \over f(x_i)}
\det \Big ( \begin{array}{ll}
(p_j(x_i))_{\substack{0 \le i < n \\ 0 \le j < n-1}} &
(r_{n-1}^{(1)}(x_i))_{0 \le i < n} \end{array} \Big ) \nonumber \\
D^{\text{odd}(M)}(x_0,\dots,x_{n-1}) & \propto
\prod_{i=1}^{n-1} {g(x_i) \over f(x_i)}
\det \Big (  \begin{array}{ll}
(p_j(x_i))_{\substack{0 \le i < n \\ 0 \le j < n-1}} &
(r_{n-1}^{(2)}(x_i))_{0 \le i < n}  \end{array} \Big ) \nonumber \\
D^{\text{alt}(M)}(x_0,\dots,x_{n-1}) & \propto
\prod_{i=1}^{n-1} {g(x_i) \over f(x_i)}
\det \Big ( \begin{array}{ll}
(p_j(x_i))_{\substack{0 \le i < n \\ 0 \le j < n-1}} &
(r_{n-1}^{(3)}(x_i))_{0 \le i < n} 
\end{array} \Big )
\quad (n \: \text{even})
 \nonumber \\
D^{\text{alt}(M)}(x_0,\dots,x_{n-1}) & \propto
\prod_{i=1}^{n-1} {g(x_i) \over f(x_i)}
\det \Big ( 
\begin{array}{ll}
(p_j(x_i))_{\substack{0 \le i < n \\ 0 \le j < n-2}} &
(r_{n-1}^{(4)}(x_i))_{0 \le i < n}  \end{array} \Big )
\quad (n \: \text{odd}),
\end{align}
while for $M = \fOE_n(f) \cup \fOE_{n+1}(f)$
\begin{align}
D^{\text{even}(M)}(x_0,\dots,x_{n-1}) & \propto
\prod_{i=1}^{n-1} {g(x_i) \over f(x_i)}
\det \Big (p_j(x_i) \Big )_{0 \le i,j < n} \nonumber \\
D^{\text{odd}(M)}(x_0,\dots,x_{n-1}) & \propto
\prod_{i=1}^{n-1} {g(x_i) \over f(x_i)}
\det \Big ( \begin{array}{lll}
(p_j(x_i))_{\substack{0 \le i < n \\ 0 \le j < n-2}} &
(r_{n-2}^{(4)}(x_i))_{0 \le i < n} & (r_{n-1}^{(5)})_{0 \le i < n} 
\end{array} \Big ).
\end{align}
In each case, setting 
$\eta_j(x)/\int_{-\infty}^\infty f(x) p_i(x)
\eta_j(x) \, dx$ equal to $g(x)/f(x)$ times the function
in column $j$ and $q_i(x) = p_i(x)$ we have that
$$
\int_{-\infty}^\infty f(x) q_i(x) \eta_j(x) \, dx = \delta_{i,j}
\quad (i,j=0,\dots,n-1),
$$
which is the desired biorthogonality property. Hence substitution of
these values into (\ref{S8}) gives the $k$-point distribution in each case.

In particular with $M = \fOE_n(f) \cup \fOE_{n+1}(f)$ and $f$ one of the
classical weights in (\ref{4.1a}) we read off that
$$
\rho_k^{{\rm even}(M)}(x_0,\dots,x_{k-1}) =
\det \Big ( 
(g(x_i) g(x_j))^{1/2}
\sum_{l=0}^{n-1} {p_l(x_i) p_l(x_j) \over (p_l,p_l)_2}
\Big )_{0\le i,j < k}.
$$
This is the well known expression for $\rho_k$ in $\fUE(g)$, and thus is in
keeping with the result of Theorem \ref{thm:OEOE.2}, giving
even$(\fOE_n(f) \cup \fOE_{n+1}(f)) = \fUE_n(g)$ for each
of the pairs $(f,g)$ in (\ref{4.1a}). Furthermore, the Christoffel-Darboux
formula evaluates the sum as
\begin{equation}\label{7.11'}
S_2(x,y) := (g(x)g(y))^{1/2}
\sum_{l=0}^{n-1} {p_l(x) p_l(y) \over (p_l,p_l)_2} =
{ (g(x) g(y) )^{1/2}
 \over (p_{n-1}, p_{n-1})_2}
{p_n(x) p_{n-1}(y) - p_{n-1}(x) p_n(y) \over x - y}
\end{equation}

\section{Distribution functions for alternate eigenvalues in a single
$\fOE_n$}

The $k$-point distribution function for the alternate eigenvalues in a single
$\fOE_n$ has a different structure to $\rho_{k}$ for the superimposed
$\fOE_n$ spectra. The cases $n$ even and $n$ odd must be treated separately.

\subsection*{$n$ even}
Consider first even$(\fOE_n(f))$ with $n$ even. From the manipulations
sketched in the proof of Theorem \ref{thm3.7} we have that the PDF of
this ensemble is given by
\begin{equation}\label{2.2}
{1 \over C} \prod_{l=0}^{n/2 - 1} f(x_{2l})
\det \left ( \begin{array}{c} x_{2i}^j \\ \int_{x_{2i}}^\infty t^j
f(t) \, dt \end{array} \right )_{0\le i < n \atop
0 \le j < 2n}
\end{equation}
To perform the integration required by (\ref{5.1}) we introduce the skew
inner product
\begin{eqnarray}\label{2.5'}
\langle u | v \rangle_1 & := & {1 \over 2} \int_{-\infty}^\infty dx \,
f(x) \Big ( u(x) \int_x^\infty dy \, f(y) v(y) - v(x)
\int_x^\infty dy \, f(y) u(y) \Big ) \nonumber  \\
& = & {1 \over 2} \int_{-\infty}^\infty dx \, f(x) u(x)
 \int_{-\infty}^\infty dy \, v(y) {\rm sgn}(y-x),
\end{eqnarray}
together with a corresponding family of monic skew orthogonal
polynomials $\{R_i(x)\}_{i=0,1,\dots}$ which are defined so that
\begin{equation}\label{2.5}
\langle R_{2i} | R_{2j+1} \rangle_1 
= - \langle R_{2j+1} | R_{2i} \rangle_1 =
r_j \delta_{i,j}, \qquad
\langle R_{2i} | R_{2j} \rangle_1 
= \langle R_{2i+1} | R_{2j+1} \rangle_1 = 0.
\end{equation}
Note that the skew orthogonality property still holds if we make the
replacement
\begin{equation}\label{2.6}
R_{2i+1}(x) \mapsto R_{2i+1}(x) + \gamma_{2i} R_{2i}(x)
\end{equation}
for arbitrary $\gamma_{2i}$. However a Gram-Schmidt type construction
shows $\{R_i(x)\}_{i=0,1,\dots}$ is unique up to this
transformation.

We will first express (\ref{2.2}) as a quaternion determinant involving
$\{ R_i(x) \}_{i=0,1,\dots}$ and then show how the property (\ref{2.5})
can be used to perform the integrations. This requires
the definition of a quaternion
determinant. We regard a quaternion as a $2 \times 2$ matrix, and a
quaternion matrix as a matrix with quaternion elements. With $n$ even and
$$
Z_n := {\bf 1}_{n/2} \otimes \left (
\begin{array}{cc} 0 & -1 \\ 1 & 0 \end{array} \right ),
$$
a $n/2 \times n/2$ quaternion matrix $Q$ is said to be self dual if
$$
Q^D := Z_n Q^T Z_n^{-1} = Q.
$$
In terms of its $2 \times 2$ sub-blocks this means that the quaternion
element in position $(kj)$ is related to the element in position
$(jk)$, $j < k$  by
$$
q_{kj} = \left ( \begin{array}{cc} d & -b \\
-c & a \end{array} \right ) \quad {\rm for} \quad
q_{jk} =  \left ( \begin{array}{cc} a & b \\
c & d \end{array} \right ).
$$
Now for a self dual quaternion matrix the determinant, to be denoted qdet,
is defined by \cite{Dy70}
\begin{equation}\label{2.a1}
{\rm qdet} \, Q =
\sum_{P \in S_{n/2}} (-1)^{n/2 - l}
\prod_l (q_{ab} q_{bc} \cdots q_{da})^{(0)}
\end{equation}
where the superscript $(0)$ denotes the operation ${1 \over 2}$Tr,
$P$ is any permutation of the indicies $(1,\dots, n/2)$ consisting of
$l$ exclusive cycles of the form $(a \to b \to c \to \cdots d \to a)$ and
$(-1)^{n/2 - l}$ is the parity of $P$. Furthermore, qdet$\,Q$ is related to
the Pfaffian via the formula \cite{Dy70}
$$
{\rm qdet} \, Q = {\rm Pf} \, Q Z_n^{-1},
$$
which since $({\rm Pf}\, Q Z_n^{-1})^2 = \det Q$ (where here $Q$ is
regarded as an ordinary $n \times n$ matrix) implies \cite{MM91}
\begin{equation}\label{2.a2}
\det Q = {\rm qdet}( Q Q^D)
\end{equation}
assuming det$\,Q$ is positive.

\begin{prop}
With $p(x_0,x_2,\dots,x_{n-2})$ denoting the PDF~(\ref{2.2}), 
$\{R_i(x)\}_{i=0,1,\dots}$ the monic orthogonal polynomials with respect to
(\ref{2.5'}) and $\{r_i\}_{i=0,1,\dots}$ the
corresponding normalizations 
we can write
\begin{equation}\label{2.7}
p(x_0,x_2,\dots,x_{n-2}) = {1 \over C}
\prod_{k=0}^{n/2 - 1} (2 r_k) \,
{\rm qdet} \Big (T(x_{2j}, x_{2k}) \Big )_{0 \le j,k < n/2}
\end{equation}
where
\begin{eqnarray}
T(x,y) & := & \sum_{k=0}^{n/2 - 1} {1 \over 2 r_k}
\Big ( {\chi}_k(y)
\chi_k^D(x) \Big )^T = \left ( \begin{array}{cc} S(x,y) & I(x,y) \\
D(x,y) & S(y,x) \end{array} \right ) \nonumber \\
\chi_k(x) & := & \left ( \begin{array}{cc}
f(x) R_{2k}(x) & f(x) R_{2k+1}(x) \\
\int_x^\infty f(t) R_{2k}(t) \, dt &
\int_x^\infty f(t) R_{2k+1}(t) \end{array} \right ) \nonumber \\
S(x,y) & = & \sum_{k=0}^{N/2 - 1} {f(y) \over 2 r_k}
\Big ( R_{2k}(y) \int_x^\infty f(t) R_{2k+1}(t) \, dt -
R_{2k+1}(y)  \int_x^\infty f(t) R_{2k}(t) \, dt \Big ) \nonumber \\
I(x,y) & = & - \int_x^y S(x,y') \, dy' \nonumber \\
D(x,y) & = & {\partial \over \partial y} S(x,y) \label{2.s}
\end{eqnarray}
\end{prop}

\begin{proof}
Because the polynomials $\{R_k(x)\}_{k=0,1,\dots}$ are monic we can add
multiples of columns in (\ref{2.2}) to obtain
\begin{eqnarray}\label{2.b}
\lefteqn{
p(x_0,x_2,\dots,x_{n-2})  =
{1 \over C} \det \left ( \begin{array}{c} f(x_{2j}) R_{k-1}(x_{2j}) \\
\int_{x_{2j}}^\infty f(t) R_{k}(t) \, dt \end{array}
\right )_{0 \le j < n/2 \atop 0 \le k < n}
 =  {1 \over C}
\prod_{k=0}^{n/2 - 1} (2r_k)} \nonumber \\&& \times
\det \left ( \begin{array}{cc}
f(x_{2j}) (2 r_{k})^{-1/2} R_{2k}(x_{2j}) &
f(x_{2j})  (2r_{k})^{-1/2} R_{2k+1}(x_{2j}) \\[.1cm]
(2r_{k})^{-1/2} \int_{x_{2j}}^\infty f(t) R_{2k}(t) \, dt
&(2 r_{k})^{-1/2} \int_{x_{2j}}^\infty f(t) R_{2k+1}(t) \, dt
\end{array} \right )_{0 \le j,k < n/2}
\end{eqnarray}
Application of (\ref{2.a2}) and the formula qdet$\, A$$\,=\,$qdet$
\, A^T$ gives the formula (\ref{2.7}) with $S(x,y)$ as specified and
formulas for $I(x,y)$ and $D(x,y)$ which are easily seen to be expressible
in terms of $S(x,y)$ as stated.
\end{proof} 

A special feature of $T(x,y)$, which follows from its definition in
(\ref{2.s}) in terms of $\chi_k(y) \chi_k^D(x)$ and the skew
orthogonality of $\{R_k(x)\}_{k=0,1,\dots}$ with respect to
(\ref{2.5'}), is the integration formulas
\begin{eqnarray}\label{10.2}
\int_{-\infty}^\infty T(x,x) \, dx & = & N/2 \nonumber \\
\int_{-\infty}^\infty T(x,y) T(y,z) \, dy & = & T(x,z)
\end{eqnarray}
As a consequence of (\ref{10.2}) and the quaternion formula
(\ref{2.7}) the integrations required to compute (\ref{5.1}) can
be carried out. Thus with (\ref{10.2}) holding it is generally true that
\cite{MM91}
\begin{equation}\label{10.3}
\int_{-\infty}^\infty dx_{2m} \, {\rm qdet} \, \Big (T(x_{2i}, x_{2j})
\Big )_{0 \le i,j \le m} =
\Big ( n/2 - (m-1) \Big ) {\rm qdet}
 \, \Big (T(x_{2i}, x_{2j}) \Big )_{0 \le i,j \le m-1}
\end{equation}
Consequently we see from (\ref{2.7}) that
\begin{equation}\label{qeven}
\rho_{k}(x_0,\dots,x_{2k-2}) = {\rm qdet} \,
\Big (T(x_{2i}, x_{2j}) \Big )_{0 \le i,j < k}.
\end{equation}

If instead of considering even$(\fOE_n(f))$ we consider
odd$(\fOE_n(f))$, the above working is essentially unchanged. Thus
(\ref{2.7}) and (\ref{2.s}) hold with the replacements
\begin{equation}\label{SQ1}
\int_x^\infty \mapsto \int_{-\infty}^x \qquad \text{and}
\qquad \{x_0,x_2,\dots,x_{n-2}\} \mapsto
\{x_1,x_3,\dots,x_{n-1}\},
\end{equation}
and with this modification of $T(x,y)$ the formula (\ref{qeven}) for
$\rho_{k}$ holds with the replacements
\begin{equation}\label{SQ2}
\{x_0,x_2,\dots,x_{2k-2}\} \mapsto \{x_1,\dots,x_{2k-1}\}.
\end{equation}

We remark that the structure of (\ref{qeven}) with $T(x,y)$ given
by (\ref{2.s}) is very similar to the general expression for
$\rho_{k}$ as computed for the ensemble $\fSE_n((g/f)^2)$. First it
is necessary to introduce monic skew orthogonal polynomials
$\{Q_k(x)\}_{k=0,1,\dots}$
and corresponding normalizations $\{q_k\}_{k=0,1,\dots}$
with respect to the
skew inner product
$$
\langle u| v\rangle_4 := \int_{-\infty}^\infty dx \, 
(g(x)/f(x))^2 \Big (
u(x) v'(x) - u(x) v'(x) \Big ).
$$
We then have \cite{NW92} (see also \cite{TW98})
\begin{equation}\label{be3}
\rho_{k}(x_0, \dots, x_{k-1}) = {\rm qdet} \, 
\Big (T_4(x_i,x_j)\Big )_{0 \le
i,j \le k}
\end{equation}
where
\begin{eqnarray}\label{be4}
T_4(x,y) & := &
\left ( \begin{array}{cc} S_4(x,y) & I_4(x,y) \\
D_4(x,y) & S_4(y,x) \end{array} \right ) \nonumber \\
S_4(x,y) & = & \sum_{k=0}^{N - 1} {f(y) \over 2 q_k}
\Big ( Q_{2k}(y) {d \over dx} \Big ( f(x) Q_{2k+1}(x) \Big )  -
Q_{2k+1}(x)  {d \over dx} \Big (
 f(x) Q_{2k}(x) \Big ) \Big ) \nonumber \\
I_4(x,y) & = & - \int_x^y S_4(x,y') \, dy' \nonumber \\
D_4(x,y) & = & {\partial \over \partial y} S_4(x,y)
\end{eqnarray}

\subsection*{$n$ odd}
The PDF for the distribution even$(\fOE_n(f))$ with $n$ odd is given by
$$
{1 \over C}
\prod_{l=0}^{(n-3)/2} f(x_{2l}) \,
\det \left ( \begin{array}{c} \left ( \begin{array}{c} x_{2i}^{2j}
\\[.1cm]
\int_{x_{2j}}^\infty f(t) t^{k} \, dt \end{array} \right )_{
0 \le i < (n-1)/2 \atop 0 \le j < n} \\
{}( \int_{-\infty}^\infty w_1(t) t^{j} \, dt)_{0 \le j < n} \end{array}
\right )
$$
As in (\ref{2.b}) we can introduce the monic polynomials
$\{ R_j(x) \}_{j=0,1,\dots}$ to rewrite this as
$$
{1 \over C} \det  \left ( \begin{array}{c}
\left ( \begin{array}{c} f(x_{2i}) R_{j}(x_{2i}) \\[.1cm]
\int_{x_{2i}}^\infty f(t) R_{j}(t) \, dt \end{array}
\right )_{0 \le i < (n-1)/2 \atop 0 \le j < n}
\\ {}( \int_{-\infty}^\infty f(t) R_{j}(t) \, dt )_{j=0,\dots,n-2}
\end{array}
\right )
$$
Subtracting appropriate multiples of the last column from the columns
$0,1,\dots, n-2$ so as to eliminate the element
of the column in the final row then gives
\begin{equation}\label{f1}
{1 \over C}
\Big ( \int_{-\infty}^\infty f(t) R_{n-1}(t) \, dt \Big )
\det \left ( \begin{array}{c} f(x_{2i}) \hat{R}_{j}(x_{2i}) \\
\int_{x_{2i}}^\infty f(t) \hat{R}_{j}(t) \, dt \end{array}
\right )_{0 \le i < (n-1)/2 \atop 0 \le j < n}
\end{equation}
where
\begin{equation}\label{f2}
\hat{R}_{j}(x) := R_{j}(x) -
\bigg ( {\int_{-\infty}^\infty f(t) R_{j}(t) \, dt \over
\int_{-\infty}^\infty f(t) R_{n-1}(t) \, dt} \bigg ) R_{n-1}(x).
\end{equation}
The determinant in (\ref{f1}) is formally the same as that in (\ref{2.b}).
Thus in the case $n$ odd $p(x_0,x_2,\dots,x_{n-3})$ can be written as in
(\ref{2.7}) but with
\begin{equation}\label{f3'}
n \mapsto n -1, \quad R_i \mapsto \hat{R}_i
\end{equation}
and $C \mapsto C'$ for some normalization $C'$.

Now we can check from the definition (\ref{f2}) that for
$j=1,\dots,n-1$ the polynomials $\hat{R}_{j-1}$ satisfy the skew
orthogonality property (\ref{2.5}). This means that the integration
formula (\ref{10.3}) again applies in this modified setting and
consequently the $k$-point distribution is given by
\begin{equation}\label{rodd}
\rho_{k}(x_0,\dots,x_{2k-2}) = {\rm qdet} \,
\Big ( f^{\rm odd}(x_{2i}, x_{2j}) \Big )_{0 \le i,j < k}.
\end{equation}
where $f^{\rm odd}$ is defined as in (\ref{2.s}) but with the
replacements (\ref{f3'}). In the case of odd$(\fOE_n(f))$ the replacements
(\ref{SQ1}) and (\ref{f3'}) must be made in (\ref{qeven}) and (\ref{2.s}),
and the replacement (\ref{SQ2}) made in (\ref{rodd}).

\subsection{Summation formulas}
It has already been remarked that $\rho_{k}$ for $\fSE_n((g/f))^2)$ has the
quaternion determinant form (\ref{be3}) and (\ref{be4}).
Furthermore it is known \cite{AFNV99} that with $f$ one of the classical
weights in (\ref{4.1a}), the quantity $S_4$ in (\ref{be4}) can be
summed to give an expression independent of the skew orthogonal
polynomials associated with $g$, and dependent only on the monic
orthogonal polynomials $\{p_i(x)\}_{i=0,1,\dots}$ associated with the
weight function $g(x)$. Explicitly
\begin{equation}\label{8.19}
2 S_4(x,y) = \Big ({g(x) \over g(y)} \Big )^{1/2}
{f(y) \over f(x)} S_2(x,y) \Big |_{n \mapsto 2n}
- \gamma_{2n-1} f(y) p_{2n}(y) \int_x^\infty f(t) p_{2n-1}(t) \, dt,
\end{equation}
where $S_2$ is specified by (\ref{7.11'}) and $\gamma_{2n-1}$ by
Lemma \ref{n}. Here we will use results from \cite{AFNV99} to obtain
an analogous summation for the quantity $S(x,y)$ in (\ref{2.s}).

Suppose $n$ is even and write
$$
\Phi_j(x) := {1 \over 2} \int_{-\infty}^\infty f(t) {\rm sgn} (x-t)
R_j(t) \, dt.
$$
Then straightforward manipulation of the definition of $S(x,y)$ allows
it to be rewritten
\begin{equation}\label{si}
S(x,y) = {1 \over 2} \Big (S_1(x,y) - S_1(\infty, y) \Big )
\end{equation}
where
$$
S_1(x,y)  =  \sum_{k=0}^{n/2-1} {f(y) \over r_k}
\Big ( \Phi_{2k}(x)R_{2k
+1}(y) -
\Phi_{2k+1}(x)R_{2k}(y) \Big )
$$
The quantity $S_1(x,y)$ occurs in the quaternion determinant formula for
$k$-point distribution of
$\fOE_n(f)$. With $f$ one of the classical forms  (\ref{4.1a})
it can be summed to give  \cite{AFNV99}
\begin{equation}\label{s11}
S_1(x,y) = \Big ({g(x) \over g(y)} \Big )^{1/2}
{f(y) \over f(x)} S_2(x,y) \Big |_{n \mapsto n-1}
+ \gamma_{n-2} f(y) p_{n-1}(y) 
{1 \over 2} \int_{-\infty}^\infty 
{\rm sgn}(x-t) f(t) p_{n-2}(t) \, dt.
\end{equation}
 From this it follows
$$
S_1(\infty, y) =  \gamma_{n-2} f(y) p_{n-1}(y) 
{1 \over 2} \int_{-\infty}^\infty 
{\rm sgn}(x-t) f(t) p_{n-2}(t) \, dt
$$
and so by (\ref{si}) we can evaluate $S(x,y)$.

\begin{prop}
For $(f,g)$ a classical pair (\ref{4.1a}), 
$\{p_j(x)\}_{j=0,1,\dots}$ monic orthogonal polynomials
with respect to the weight function $g(x)$, and $n$ even the quantity
$S(x,y)$ in (\ref{2.s}) has the evaluation
\begin{equation}\label{s22}
2S(x,y) = \Big ({g(x) \over g(y)} \Big )^{1/2}
{f(y) \over f(x)} S_2(x,y) \Big |_{n \mapsto n-1}
- \gamma_{n-2} f(y) p_{n-1}(y) 
\int_{x}^\infty 
f(t) p_{n-2}(t) \, dt
\end{equation}
(c.f.~(\ref{8.19})).
\end{prop}
This summation fully determines even$(\fOE_n(f))$
with $n$ even. For odd$(\fOE_n(f))$, $n$ even, the
prescription (\ref{SQ1})  says the replacement
$\int_x^\infty \mapsto \int_{-\infty}^x$ should be made in (\ref{s22}).

It remains to consider the case $n$ odd. 
Consider first even$(\fOE_n(f))$. In fact the formulas in
\cite{AFNV99} giving the analogous formula to (\ref{s11}) for
$n$ odd allows us to deduce that the summation (\ref{s22}) remains
valid for $n$ odd. For $n$ odd comparison of (\ref{s22}) and (\ref{8.19})
shows
\begin{equation}
S(x,y) = S_4(x,y) \Big |_{n \mapsto (n-1)/2},
\end{equation}
which because of the formulas (\ref{2.s}) (with $n \mapsto n-1$),
(\ref{qeven}), and (\ref{be3}), (\ref{be4}) implies
\begin{equation}
\rho_{k}^{\text{even}(\fOE_{2n+1}(f))}
(x_0,x_2,\dots,x_{2k-2}) =
\rho_{k}^{\fSE_n((g/f)^2)}(x_0,x_2,\dots,x_{2k-2}).
\end{equation}
This is equivalent to the second statement of (\ref{3.85}), which we 
already know from
Theorem \ref{thm4.1} is valid for the pairs $(f,g)$ in (\ref{4.1a}).
In the case of odd$(\fOE_n(f))$ with $n$ odd, again the prescription
(\ref{SQ1}) says we simply make the replacement
$\int_x^\infty \mapsto \int_{-\infty}^x$ in (\ref{s22}).

\section*{Acknowledgements}
PJF thanks J.~Baik for drawing his attention to the conjecture noted in
the paragraph above the paragraph containing (\ref{0.8}), and acknowledges
the Australian Research Council for financial support.

\end{document}